\documentclass[12pt]{article}
\usepackage{amsfonts,amsthm,amsmath,amssymb,upgreek,bm,mathtools}
\usepackage[bookmarks = false, linktocpage=true]{hyperref}
\usepackage[paper=letterpaper,margin=.85in]{geometry}
\usepackage{graphicx}
\usepackage{units}
\usepackage{float}
\usepackage[export]{adjustbox}
\usepackage{bbold}
\usepackage{xcolor}
\usepackage[shortlabels]{enumitem}
\usepackage{dsfont}
\usepackage[mathscr]{euscript}
\usepackage[normalem]{ulem}
\usepackage{braket}
\usepackage{slashed}
\usepackage{comment}

\newcommand{\be}{\begin{equation}}
\newcommand{\ee}{\end{equation}}
\renewcommand{\tilde}{\widetilde}

\newcommand{\Z}{\mathbb{Z}}

\newcommand{\N}{\mathbb{N}}

\renewcommand{\H}{\mathcal{H}}

\numberwithin{equation}{section}

\def\tr{\text{Tr}}

\usepackage{color}
\definecolor{darkblue}{rgb}{0,0,0.6}
\definecolor{purple}{rgb}{0.4,.2,0.7}
\definecolor{darkgreen}{rgb}{0,0.5,0}

\begin{document}
\thispagestyle{empty}

\vspace*{2.5cm}
\begin{center}

{\bf {\LARGE More on Torus Wormholes in 3d Gravity}}

\begin{center}

\vspace{1cm}

{\bf Cynthia Yan}\\
  \bigskip \rm

\bigskip 

Stanford Institute for Theoretical Physics,\\Stanford University, Stanford, CA 94305

\rm
  \end{center}

\vspace{2.5cm}
{\bf Abstract}
\end{center}
\begin{quotation}
\noindent

We study further the duality between semiclassical AdS$_3$ and formal CFT$_2$ ensembles. First, we study torus wormholes (Maldacena-Maoz wormholes with two torus boundaries) with one insertion or two insertions on each boundary and find that they give non-decaying contribution to the product of two torus one-point or two-point functions at late-time. Second, we study the $\Z_2$ quotients of a torus wormhole such that the outcome has one boundary. We identify quotients that give non-decaying contributions to the torus two-point function at late-time.

We comment on reflection (R) or time-reversal (T) symmetry v.s.~the combination RT that is a symmetry of any relativistic field theory. RT symmetry itself implies that to the extent that a relativistic quantum field theory exhibits random matrix statistics it should be of the GOE type for bosonic states and of the GSE type for fermionic states. We discuss related implications of these symmetries for wormholes.

\end{quotation}

 \newcommand{\psltwor}{\text{PSL}(2,\mathbb{R})}
  \newcommand{\psltwoc}{\text{PSL}(2,\mathbb{C})}
 \newcommand{\volhtwo}{\text{vol}(H^2)}
 \newcommand{\osp}{\text{OSp}(1|2)/\mathbb{Z}_2}

\setcounter{page}{0}
\setcounter{tocdepth}{3}
\setcounter{footnote}{0}
\newpage

\setcounter{page}{2}
\tableofcontents
\pagebreak

\section{Introduction}

It was conjectured that there is an approximate duality between semiclassical 3d gravity and an ensemble of formal CFT$_2$s with large central charge $c$ and a sparse low-energy spectrum \cite{Chandra:2022bqq, Belin:2020hea, Collier:2019weq} \footnote{This is not a true microscopic ensemble. For a free toy model of CFT$_2$ ensemble microscopically defined see \cite{Maloney:2020nni, Afkhami-Jeddi:2020ezh}}. This is analogous to the duality between Jackiw-Teitelboim (JT) gravity \cite{Teitelboim, Jackiw, AlmheiriPolchinski} and random matrix theory \cite{Saad:2019lba, Stanford:2019vob}. These dualities are both different from earlier examples of AdS/CFT and their boundary ensembles satisfy the Eigenstate Thermalization Hypothesis (ETH) \cite{ethSrednicki, ethDeutsch}.

This proposed duality provides a playground to study various quantities. One thing is a BTZ black hole \cite{Banados:1992wn} in AdS$_3$ with light operator insertions. Here light and heavy are respectively distinguished by whether the operator dimension $(h,\bar{h})$ is below or above the black hole threshold $\frac{c}{24}$ where $c=\frac{3}{2G}$ is the central charge. In particular, we study averaged one-point and two-point correlators on a torus on the CFT$_2$ ensemble side. Their variances involve a torus wormhole on the 3d gravity side. More precisely, the averaged product of two torus correlators are dual to a Maldacena-Maoz wormhole \cite{Maldacena:2004rf} with insertions. A Maldacena-Maoz wormhole has two asymptotic boundaries with identical topology and constant moduli. The boundaries are Riemann surfaces, with the simplest options Riemann spheres or tori. In order for a Riemann sphere to be stable, there need to be at least three insertions. For torus, at least one insertion. If we take the operator dimension to zero for a torus wormhole, we can connect back to \cite{Cotler:2020ugk, Cotler:2020hgz} which calculates the torus wormhole partition function in 3d pure gravity \cite{Maloney:2007ud}.



\begin{figure}[H]
\centering
\includegraphics[width=\textwidth]{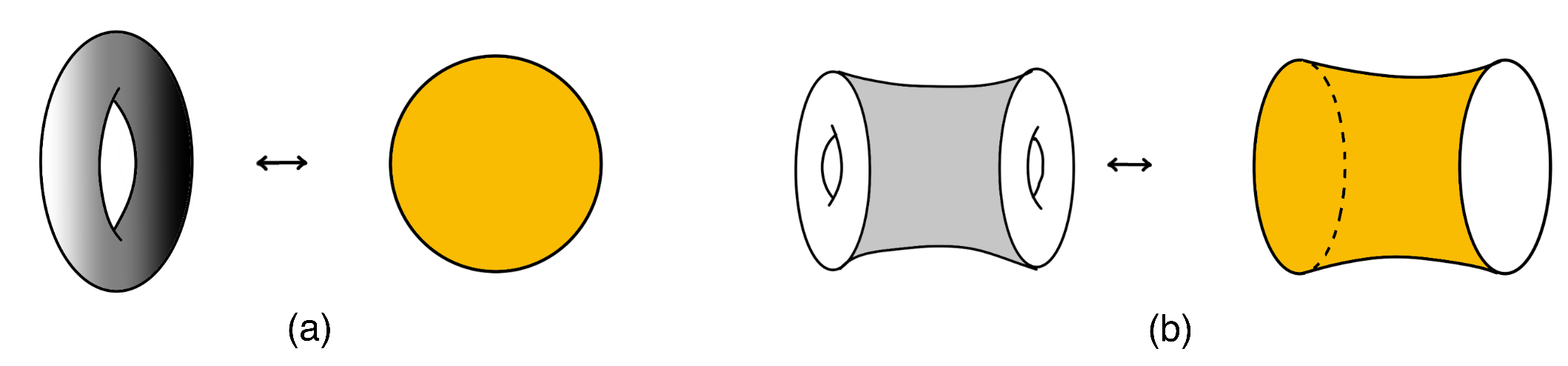}
\caption{(a) Analogy between a 3d solid torus and a 2d disk (b) Analogy between a 3d torus wormhole and a 2d cylinder}
\label{3dto2d}
\end{figure}

The BTZ black hole can be reduced to JT gravity by Kaluza-Klein mechanism \cite{Ghosh:2019rcj, Maxfield:2020ale}. In this paper we study 3d gravity using methods inspired by 2d gravity. In particular, there is an analogy between a solid torus in 3d and a disk in 2d as shown in figure \ref{3dto2d}(a). This can be extended to the analogy between a torus wormhole in 3d and a cylinder (or double-trumpet) in 2d as shown in figure \ref{3dto2d}(b). 

\begin{figure}[H]
\centering
\includegraphics[width=\textwidth]{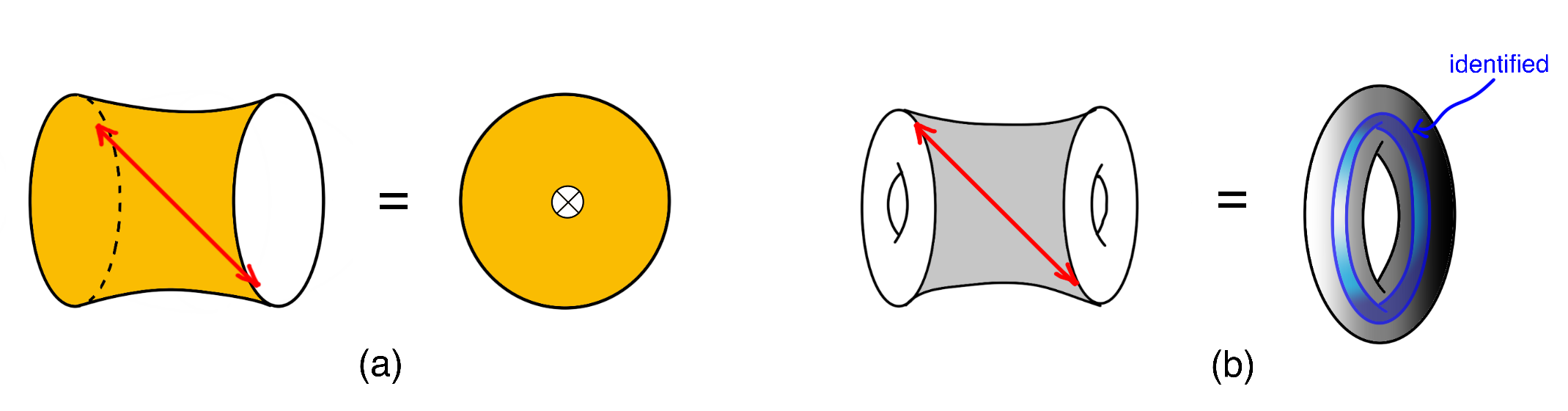}
\caption{(a) $\Z_2$ quotient of a cylinder in 2d (b) $\Z_2$ quotients of a torus wormhole in 3d}
\label{3dto2dcc1}
\end{figure}

We can take the analogy further by taking a $\Z_2$ quotient of a torus wormhole in 3d and compare that to a $\Z_2$ quotient of a cylinder in 2d. In 2d, identifying antipodal points on a cylinder gives a disk with a crosscap inserted as shown in figure \ref{3dto2dcc1}(a). Similarly in 3d, we identify the torus wormhole as shown in figure \ref{3dto2dcc1}(b) and get a solid torus with a thinner solid torus carved out in the middle. Every point on the blue torus is identified with another point on that blue torus. We find that there are two ways of doing the $\Z_2$ quotient; one orientable and one non-orientable. Each configuration gives a non-decaying contribution to torus two-point function reminiscent to that a disk with a crosscap gives a non-decaying contribution to the thermal two-point function in 1d \cite{Yan:2022nod}.

\noindent We can understand these two contributions using Random Matrix Theory (RMT) from the boundary side perspective. In particular, the orientable geometry comes from RT symmetry contained in the proposed formal CFT$_2$ ensemble. In 2d, CPT symmetry is equivalent to reflection+time reversal (RT). While the non-orientable geometry arises from adding time reversal (T) symmetry to the formal CFT$_2$ ensemble. Noticing this RT symmetry has bigger consequences, one is that it tells us that the torus wormhole partition function calculated in \cite{Cotler:2020ugk, Cotler:2020hgz} should be multiplied by a factor of two. The other is that we notice that actually any relativistic quantum field theory with random matrix statistics has energy eigenvalue distribution a GOE ensemble for bosonic states and a GSE ensemble for fermionic states. 

We show the 3d calculations in the main text and the parallel 2d calculations in Appendix \ref{2dappendix}. In section \ref{onepoint}, we consider one insertion on each boundary torus. First in section \ref{CJlimit}, we find that if we take the mass of the insertions to zero, we get a result similar to the partition function found in \cite{Cotler:2020ugk, Cotler:2020hgz} but with a factor off. Second in section \ref{1ptlatetime}, we find that if we analytically continue the two boundary tori, the average product of one-point functions do not decay over time. In section \ref{twopoint}, we consider two insertions on each boundary torus. First in section \ref{2ptlatetime}, we find that if we analytically continue the location of one of the insertions on each boundary, the average product of two-point functions do not decay over time. Second in section \ref{3dcrosscap} we discuss the $\Z_2$ quotients of a torus wormhole and their contributions to torus two-point function both from a bulk perspective and from a boundary perspective, and examine the effect of RT and T symmetry. Finally, in section \ref{discussion} we look again at the partition function found by \cite{Cotler:2020ugk, Cotler:2020hgz} and also examine a generic relativistic quantum field theory with random matrix statistics.

\subsection*{Review: CFT$_2$ ensemble}
\label{cftreview}
We review some key properties of unitary compact CFT$_2$ following the presentation of \cite{Collier:2019weq} and state the conjectures of \cite{Chandra:2022bqq} that we will also assume in this paper. A CFT$_2$ is defined by its central charge $c$ and a list of primary operators with known scaling dimensions $(h,\bar{h})$ and operator product expansion (OPE) coefficients $c_{ijk}$. We reparametrize in terms of ``Liouville parameters''
\be
c=1+6Q^2=1+6(b+\frac{1}{b})^2\quad\quad h=\alpha(Q-\alpha)\quad\quad \alpha=\frac{Q}{2}+iP
\ee
Also denote the spin as $s=h-\bar{h}$. Viewed as an ensemble, CFT$_2$'s have some universal features depending only on the central charge, we list two such features below. Both can be formulated in two ways which are equivalent to each other because of modular invariance.

\begin{enumerate}
\item
\begin{enumerate}
\item The identity operator has
\be
h_{\mathbb{1}}=0=\bar{h}_{\mathbb{1}}
\ee
\item Cardy's formula for the density of primary states
\be
\rho(h,\bar{h})\approx\exp\{4\pi(\sqrt{\frac{(c-1)h}{24}}+\sqrt{\frac{(c-1)\bar{h}}{24}})\}\quad\text{as}\quad h,\bar{h}\rightarrow\infty
\ee
\end{enumerate}
\item 
\begin{enumerate}
\item For any operator $O_i$, the OPE coefficient of two $O_i$'s and the identity is
\be
c_{ii\mathbb{1}}=1
\ee
\item Averaged value of the OPE coefficients over all primary operators is given by
\be 
\overline{|c_{abc}|^2}=C_0(h_a,h_b,h_c)C_0(\bar{h}_a,\bar{h}_b,\bar{h}_c)
\ee
\end{enumerate}
when at least one of the three operator is heavy. Here $C_0$ is given by 
\be
C_0(h_1,h_2,h_3)=C_0(P_1,P_2,P_3)=\frac{\Gamma_b(2Q)}{\sqrt{2}\Gamma_b(Q)^3}\frac{\Gamma_b(\frac{Q}{2}\pm iP_1\pm iP_2 \pm iP_3)}{\prod_{k=1}^3\Gamma_b(Q\pm 2iP_k)}\label{ceq}
\ee
where
\be
\Gamma_b(x)=\frac{\Gamma_2(x|b,b^{-1})}{\Gamma_2(Q/2|b,b^{-1})}
\ee
\end{enumerate}

From now on, following the conventions of \cite{Chandra:2022bqq}, we assume that our CFTs are holographic, i.e.  with a large central charge and a sparse spectrum of low-lying operators. We also extend 1(b) to hold above the black hole threshold and 2(b) to hold for all primaries. In addition assume
\be
\overline{c_{abc}}=0
\ee
and 
\be
\overline{c_{abc}c^*_{def}}=C_0(h_a,h_b,h_c)C_0(\bar{h}_a,\bar{h}_b,\bar{h}_c)(\delta_{ad}\delta_{be}\delta_{cf}+(-1)^{s_a+s_b+s_c}\delta_{ad}\delta_{bf}\delta_{ce}+\text{4 more terms})\label{Tensemble}
\ee
where the remaining four terms are cyclic permutations of the first two terms. In section \ref{crosscapcft}, we discuss a way to understand this ensemble from the prospective of Random Matrix Theory.

We introduce a graphical way of representing the OPE coefficients. Consider a CFT$_2$ on a Riemann sphere. Let $O_1$, $O_2$, and $O_3$ be three insertions. Instead of thinking of these as point insertions we can expand the point out to form cycles and the Riemann sphere with three insertions now look like a pair of pants. This just gives the OPE coefficient $c_{123}$.

\be
\braket{O_1O_2O_3}=\includegraphics[valign=c,width=0.15\textwidth]{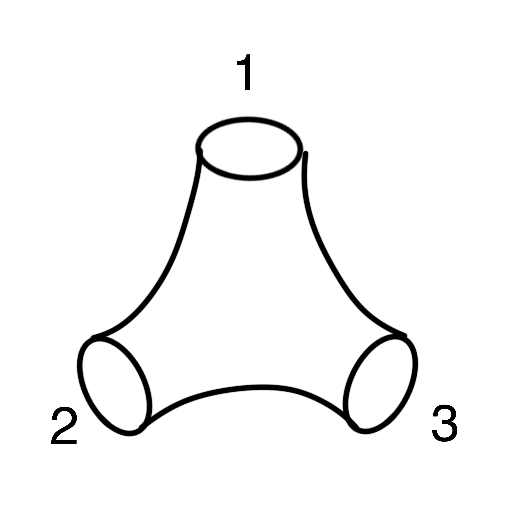}=c_{123}
\ee

\begin{figure}[H]
\centering
\includegraphics[width=0.2\textwidth]{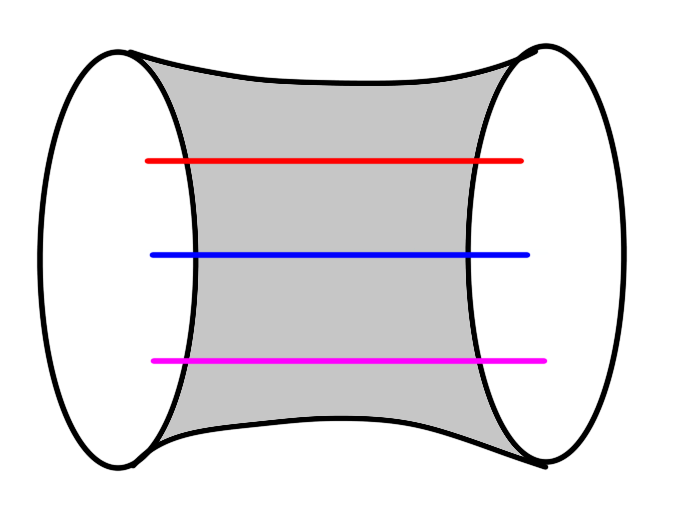}
\caption{A Maldacena-Maoz wormhole with two boundaries both Riemann spheres with three light insertions}
\label{wormhole3}
\end{figure}

\cite{Chandra:2022bqq} showed that the averaged product of two three-point-functions of two CFT$_2$'s on Riemann spheres matches the semiclassical action of the sphere wormhole shown in figure \ref{wormhole3} on the 3d gravity side. 

\noindent They are both equal to the universal asymptotic formula for OPE coefficients i.e.
\be
\overline{\braket{O_1O_2O_3}_{S^2}\braket{O_{1'}O_{2'}O_{3'}}}_{S^2}=\overline{c_{123}c_{1'2'3'}}=|C_0(h_1,h_2,h_3)|^2\label{ensembleeq}
\ee

\section{Torus one-point function wormhole}
\label{onepoint}

In this section, we focus on studying a torus wormhole with one insertion on each boundary. In section \ref{CJlimit} we take the mass of the insertions to zero and compare with partition function in \cite{Cotler:2020ugk, Cotler:2020hgz}, in section \ref{1ptlatetime} we show that the averaged product of two torus one-point functions does not decay over time. We consider the averaged product of two one-point functions.
\be
\overline{\braket{O_1}_{T^2(\tau,\bar{\tau})}\braket{O_1}_{T^2(\tau',\bar{\tau}')}}
\ee
Note that pictorially, for each torus one-point function we can think of it as a torus with a hole. Then this can be decomposed as a sum of the product of OPE coefficient $c_{1pp}$ and conformal block $|\mathcal{F}_1^{g=1}(h_p;\tau)|^2$ over primary operators $p$. The sum over descendents of each primary operator is encoded in the conformal block.
\begin{align}
\braket{O_1}_{T^2(\tau,\bar{\tau})}&=\includegraphics[valign=c,width=0.1\textwidth]{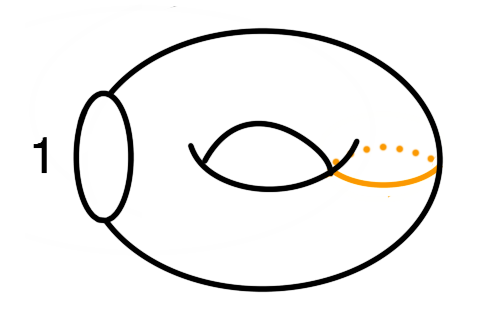}\\
&=\sum_p \includegraphics[valign=c,width=0.1\textwidth]{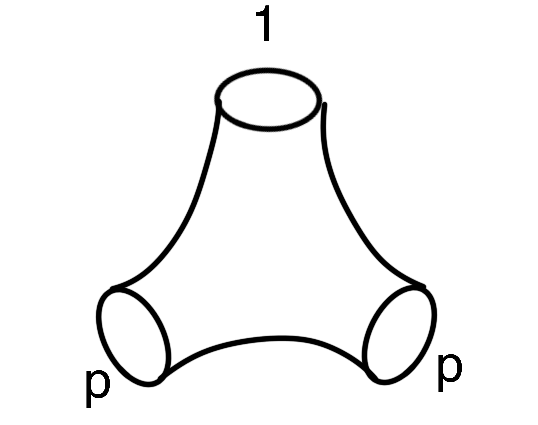}\left|\includegraphics[valign=c,width=0.1\textwidth]{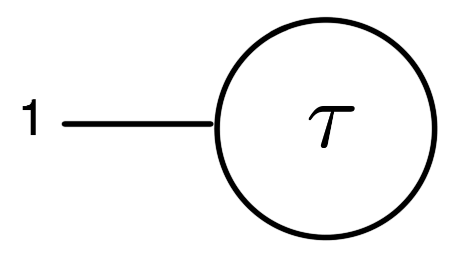}\right|^2\\
&=\sum_pc_{1pp}|\mathcal{F}_1^{g=1}(h_p;\tau)|^2\\
&=\sum_pc_{1pp}\mathcal{F}_1^{g=1}(h_p;\tau)\overline{\mathcal{F}}_1^{g=1}(\bar{h}_p;\bar{\tau})\label{1ptdiscrete}
\end{align}
The proposed ensemble of formal CFT$_2$ (\ref{ensembleeq}) predicts that the averaged product of two torus one-point functions is given by 
\begin{align}
\overline{\braket{O_1}_{T^2(\tau,\bar{\tau})}\braket{O_1}_{T^2(\tau',\bar{\tau}')}}&\approx\left(1+(-1)^{s_1}\right)\big|\int dh_p\rho_0(h_p)C_0(h_1,h_p,h_p)\mathcal{F}_1^{g=1}(h_p;\tau)\overline{\mathcal{F}}_1^{g=1}(h_p;-\tau')\big|^2\label{werwer}\\
&=\left(1+(-1)^{s_1}\right)\braket{O_1}^L_{T^2(\tau,-\tau')}\braket{O_1}^L_{T^2(-\bar{\tau}',\bar{\tau})}\label{1ptresult}
\end{align}
where $\braket{O_1}^L_{T^2(\tau,-\tau')}$ and $\braket{O_1}^L_{T^2(-\bar{\tau}',\bar{\tau})}$ are torus Liouville one-point functions. For a derivation of this result see Appendix \ref{apponept}. 

It was shown in \cite{Chandra:2022bqq} that in the large $c$ limit (to one-loop order), (\ref{1ptresult}) agrees with the semiclassical limit of the 3d gravity calculation of a torus wormhole with a defect insertion. It is interesting to contemplate that the RHS of (\ref{werwer}) could actually be the exact answer for this wormhole in 3d gravity. Our understanding is that this has indeed been established as part of the work of \cite{Lorenztalk}. This motivates us to take the formula seriously enough to consider a non-semiclassical limit where $\Delta \to 0$.

\subsection{$\Delta\rightarrow0$ limit}
\label{CJlimit}
In this section, we take the mass of the insertions to zero $\Delta\rightarrow0$. There are a couple of interesting features that we want to point out about this limit. First, the expression is simple and given by
\be
\overline{\braket{O_1}_{T^2(\tau,\bar{\tau})}\braket{O_1}_{T^2(\tau',\bar{\tau}')}}=\frac{4\left(1+(-1)^{s_1}\right)Q^2}{\Delta^2\pi^2}\int dkd\bar{k}\chi_k(\tau)\chi_k(\tau')\chi_{\bar{k}}(\bar{\tau})\chi_{\bar{k}}(\bar{\tau}')
\ee
where the character is given by
\be
\chi_k(\tau)=\frac{q^{\frac{k^2}{4}}}{\eta(\tau)}=\frac{e^{\frac{i\pi\tau k^2}{2}}}{\eta(\tau)}\quad\quad k=2P=\sqrt{4h_p-Q^2}
\ee
Here $\eta(\tau)$ is the Dedekind eta function. Second, naively we would guess that in the $\Delta\rightarrow0$ limit, the averaged product of two one-point functions $\overline{\braket{O}\braket{O}}$ should reproduce the partition function predicted by Cotler-Jensen  \cite{Cotler:2020ugk, Cotler:2020hgz}. This turns out to not to be the case, and the discrepancy is simple, just a factor inside our integral over $k$ and $\bar{k}$. The same kind of discrepancy also appears in 2d because even though we take $\Delta\rightarrow0$ there is a still a sum over windings of the particle around the cylinder which contributes a divergent overall factor (see Appendix \ref{2donept} for a detailed 2d analysis).

Let us begin with our formula of the averaged product of two one-point functions
\be
\overline{\braket{O_1}_{T^2(\tau,\bar{\tau})}\braket{O_1}_{T^2(\tau',\bar{\tau}')}}\approx\left(1+(-1)^{s_1}\right)\big|\int dh_p\rho_0(h_p)C_0(h_1,h_p,h_p)\mathcal{F}_1^{g=1}(h_p;\tau)\overline{\mathcal{F}}_1^{g=1}(h_p;-\tau')\big|^2
\ee
where $C_0$ is given by (\ref{ceq}). Now we take the weight of $O_1$ to zero, i.e. $\Delta=2h_1\rightarrow0$. In this limit $O_1$ becomes the identity operator. Then the above equation becomes
\be
\overline{\braket{O_1}_{T^2(\tau,\bar{\tau})}\braket{O_1}_{T^2(\tau',\bar{\tau}')}}\approx\frac{4\left(1+(-1)^{s_1}\right)Q^2}{\Delta^2\pi^2}\left|\int dP\,\mathcal{F}_\mathbb{1}^{g=1}(h_p;\tau)\overline{\mathcal{F}}_\mathbb{1}^{g=1}(h_p;-\tau')\right|^2
\ee
for derivation see \ref{appdetailc0}.

From (\ref{1ptdiscrete}) we know that
\be
Z_{T^2}(\tau)=\sum_p\mathcal{F}_\mathbb{1}^{g=1}(h_p;\tau)\overline{\mathcal{F}}_\mathbb{1}^{g=1}(\bar{h}_p;\bar{\tau})
\ee
But we also know that
\be
Z_{T^2}(\tau)=\sum_p\chi_k(\tau)\bar{\chi}_k(\bar{\tau}) 
\ee
so
\be
\mathcal{F}_\mathbb{1}^{g=1}(h_p;\tau)\overline{\mathcal{F}}_\mathbb{1}^{g=1}(\bar{h}_p;\bar{\tau})=\chi_k(\tau)\bar{\chi}_k(\bar{\tau}) 
\ee
and the averaged product of two torus one-point function in the limit $\Delta\rightarrow0$ is given by
\be
\overline{\braket{O_1}_{T^2(\tau,\bar{\tau})}\braket{O_1}_{T^2(\tau',\bar{\tau}')}}=\frac{4\left(1+(-1)^{s_1}\right)Q^2}{\Delta^2\pi^2}\int dkd\bar{k}\,\chi_k(\tau)\chi_k(\tau')\chi_{\bar{k}}(\bar{\tau})\chi_{\bar{k}}(\bar{\tau}')\label{1ptCJ}
\ee
On the other hand, the partition function given in Cotler\&Jensen \cite{Cotler:2020ugk, Cotler:2020hgz} is
\be
Z_{T^2\times I}(\tau_1,\bar{\tau}_1,\tau_2,\bar{\tau}_2)=\sum_{\gamma\in\mathrm{PSL}(2;\Z)}\tilde{Z}(\tau_1,\bar{\tau}_1,\gamma\tau_2,\gamma\bar{\tau}_2)\label{CJsum}
\ee
where
\be
\tilde{Z}(\tau_1,\bar{\tau}_1,\tau_2,\bar{\tau}_2)=\sqrt{\mathrm{Im}(\tau_1)\mathrm{Im}(\tau_2)}\int_0^\infty dkd\bar{k}\chi_k(\tau_1)\chi_k(\tau_2)\bar{\chi}_{\bar{k}}(\bar{\tau}_1)\bar{\chi}_{\bar{k}}(\bar{\tau}_2)k\bar{k}
\ee
so the $\lim_{\Delta\rightarrow0}\overline{\braket{O}\braket{O}}$ differ by a factor $k\bar{k}$ inside the integral compared to the partition function $\tilde{Z}$ before the sum over $\mathrm{PSL}(2,\mathbb{Z})$. However, not like in 2d, in 3d we do not know a precise way to calculate this factor yet.

We can also compare the expressions after integration which are
\be
\overline{\braket{O_1}_{T^2(\tau,\bar{\tau})}\braket{O_1}_{T^2(\tau',\bar{\tau}')}}=\frac{4\left(1+(-1)^{s_1}\right)Q^2}{\Delta^2\pi^2}\frac{1}{|\eta(\tau_1)|^2|\eta(\tau_2)|^2}\frac{1}{|\tau_1+\tau_2|}
\ee
v.s.
\be
\tilde{Z}(\tau_1,\bar{\tau}_1,\tau_2,\bar{\tau}_2)=\frac{\sqrt{\mathrm{Im}(\tau_1)\mathrm{Im}(\tau_2)}}{2\pi^2|\eta(\tau_1)|^2|\eta(\tau_2)|^2}\frac{1}{|\tau_1+\tau_2|^2}
\ee

\subsection{Lorentzian torus limit}
\label{1ptlatetime}
In this section, we examine the late-time behavior of the product of two torus one-point functions. In 2d, the simplest quantity to probe the late-time behavior is the spectral form factor $\braket{Z(\beta+it)Z(\beta-it)}$ \cite{Saad:2018bqo}. This is two thermal partition functions, one with $\beta$ analytically continued to $\beta+it$ and the other to $\beta-it$. We can do a similar analytical continuation in 3d. 

Writing the torus one-point function in operator form, we get
\be
\braket{O_1}_{T^2(\tau,\bar{\tau})}=\mathrm{tr}(O_1e^{-\beta H+isP})=\mathrm{tr}(O_1e^{i\tau(h-c/24)}e^{-i\bar{\tau}(\bar{h}-c/24)})
\ee
where $\tau=i\beta+s$ and $\bar{\tau}=-i\beta+s$ are complex conjugates since the torus $T^2$ is a Euclidean torus. Also by definition $H=h+\bar{h}-c/12$ and $P=h-\bar{h}$. 

The above is one boundary of the Maldacena-Maoz wormhole. The other side $T^2(\tau',\bar{\tau}')$ is another torus that look like the reflection of $T^2(\tau,\bar{\tau})$. Thus $\tau'=i\beta-s$ and $\bar{\tau}'=-i\beta-s$. Then the torus one-point function is 
\be
\braket{O_1}_{T^2(\tau',\bar{\tau}')}=\mathrm{tr}(O_1e^{-\beta H-isP})
\ee
For simplicity, we take $s=0$ so the product of the two torus one-point function becomes
\be
\braket{O_1}_{T^2(\tau,\bar{\tau})}\braket{O_1}_{T^2(\tau',\bar{\tau}')}=\mathrm{tr}(O_1e^{-\beta H})\mathrm{tr}(O_1e^{-\beta H})
\ee
Now we can do analytical continuation similar to $\braket{Z(\beta+it)Z(\beta-it)}$ \cite{Saad:2018bqo}, and make the two tori into Lorentzian tori by taking $\beta\mapsto\beta+it$ for $T^2(\tau,\bar{\tau})$ while taking $\beta\mapsto\beta-it$ for $T^2(\tau,\bar{\tau})$. But this is just saying now
\begin{align}
\tau&=i\beta-t\quad\quad\tau'=i\beta+t\\
\bar{\tau}&=-i\beta+t\quad\quad\bar{\tau}'=-i\beta-t
\end{align}
so we are calculating the quantity
\be
\braket{O_1}_{T^2(\tau,\bar{\tau})}\braket{O_1}_{T^2(\tau',\bar{\tau}')}=\braket{O_1}_{T^2(i\beta-t,-i\beta+t)}\braket{O_1}_{T^2(i\beta+t,-i\beta-t)}=\mathrm{tr}(O_1e^{-(\beta+it) H})\mathrm{tr}(O_1e^{-(\beta-it) H})
\ee
(\ref{1ptresult}) then becomes
\be
\overline{\braket{O_1}_{T^2(\tau,\bar{\tau})}\braket{O_1}_{T^2(\tau',\bar{\tau}')}}=\left(1+(-1)^{s_1}\right)\braket{O_1}^L_{T^2(i\beta-t,-i\beta-t)}\braket{O_1}^L_{T^2(i\beta+t,-i\beta+t)}
\ee
But now we can see that the Liouville one-point functions on the right-hand-side are on Euclidean tori instead of Lorentzian tori. Another way to present the answer is
\be
\overline{\mathrm{tr}(O_1e^{-(\beta+it) H})\mathrm{tr}(O_1e^{-(\beta-it) H})}=\left(1+(-1)^{s_1}\right)\mathrm{tr}_L(O_1e^{-\beta H-itP})\mathrm{tr}_L(O_1e^{-\beta H+itP})
\ee
On the LHS we start with an average over the formal ensemble of CFTs of a quantity similar to the spectral form factor, but with an operator inserted in each factor. The RHS is the answer for this ensemble average, or equivalently, it is the wormhole contribution in 3d gravity -- either way it boils down to a computation in Liouville theory involving an operator insertion with the same dimension. The key point is that on the RHS, the large time parameter $t$ multiplies the momentum operator $P$, not the Hamiltonian $H$. Now, $P$ is quantized so the RHS is periodic in time with a short period, and in particular does not decay for large $t$.

\section{Torus two-point function wormholes}
\label{twopoint}
In this section we focus on studying a torus wormhole with two insertions on each boundary. In section \ref{2ptlatetime}, we show that the averaged product of two torus two-point functions does not decay over time and in section \ref{3dcrosscap} we study $\Z_2$ quotients of this torus wormhole. These quotients give non-decaying contributions to the torus two-point function. We calculate these contributions from the bulk in section   \ref{crosscapgravity} and from the boundary in section \ref{crosscapcft} where we justify and extend the proposed formal CFT$_2$ ensemble (\ref{Tensemble}) using RMT. 

We consider the averaged product of two torus two-point functions
\be
\overline{\braket{O_1(v,\bar{v})O_2(0)}_{T^2(\tau,\bar{\tau})}\braket{O_1(v',\bar{v}')O_2(0)}_{T^2(\tau',\bar{\tau}')}}
\ee
Each two-point function is given by
\begin{align}
\braket{O_1(v,\bar{v})O_2(0)}_{T^2(\tau,\bar{\tau})}&=\includegraphics[valign=c,width=0.15\textwidth]{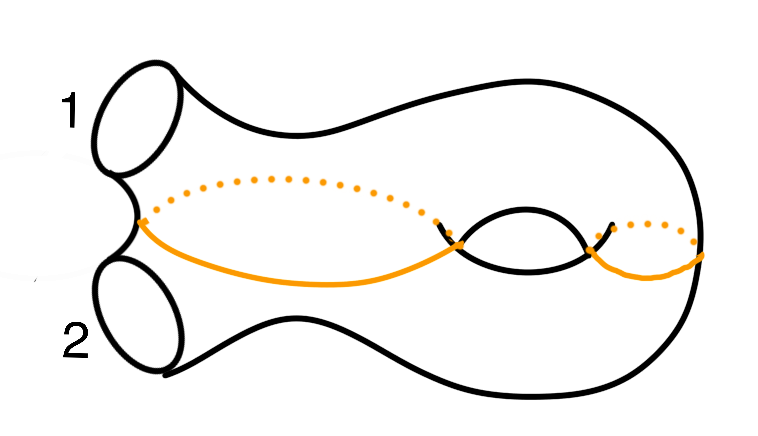}\\
&=\sum_{p,q} \includegraphics[valign=c,width=0.1\textwidth]{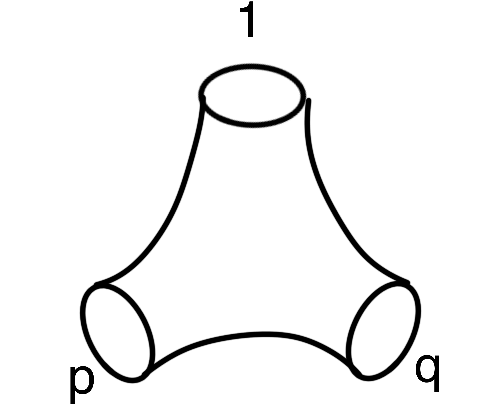}\includegraphics[valign=c,width=0.1\textwidth]{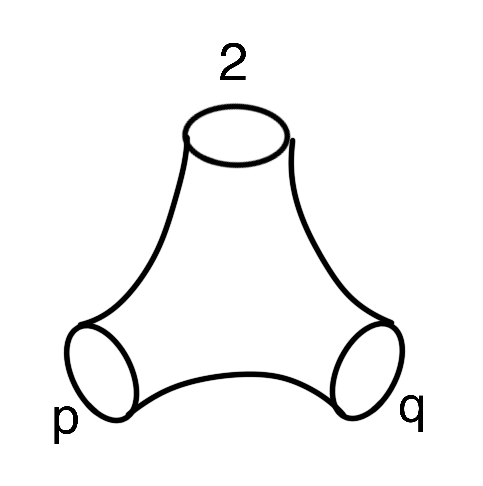}\left|\includegraphics[valign=c,width=0.1\textwidth]{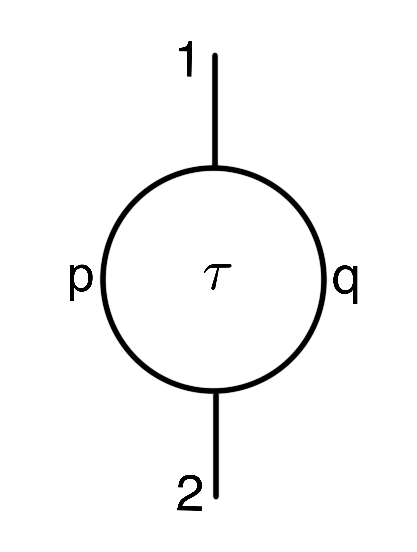}\right|^2\\
&=\sum_pc_{1pq}c_{2pq}|\mathcal{F}_{12}^{g=1}(h_p,h_q;\tau,v)|^2\\
&=\sum_pc_{1pq}c_{2pq}\mathcal{F}_{12}^{g=1}(h_p,h_q;\tau,v)\overline{\mathcal{F}}_{12}^{g=1}(\bar{h}_p,\bar{h}_q;\bar{\tau},\bar{v})
\end{align}
Thus the averaged product of two torus two-point functions is given by
\begin{align}
&\overline{\braket{O_1(v,\bar{v})O_2(0)}_{T^2(\tau,\bar{\tau})}\braket{O_1(v',\bar{v}')O_2(0)}_{T^2(\tau',\bar{\tau}')}}\nonumber\\
\approx&2\big|\int dh_p dh_q\rho_0(h_p)\rho_0(h_q)C_0(h_1,h_p,h_q)C_0(h_2,h_p,h_q)\mathcal{F}_{12}^{g=1}(h_p,h_q;\tau,v)\overline{\mathcal{F}}_{12}^{g=1}(h_p,h_q;-\tau',-v')\big|^2\\
\approx& 2\braket{O_1(v,-v')O_2(0)}^L_{T^2(\tau,-\tau')}\braket{O_1(-\bar{v}',\bar{v})O_2(0)}^L_{T^2(-\bar{\tau}',\bar{\tau})}\label{2ptresult}
\end{align}
where $\braket{O_1(v,-v')O_2(0)}^L_{T^2(\tau,-\tau')}$ and $\braket{O_1(-\bar{v}',\bar{v})O_2(0)}^L_{T^2(-\bar{\tau}',\bar{\tau})}$ are torus Liouville two-point functions. For details, see Appendix \ref{apptwopt}.

\subsection{Large time-separation limit}
\label{2ptlatetime}
In this section, we examine the late-time behavior of the product of two torus two-point functions. 
\be
\braket{O_1(v,\bar{v})O_2(0)}_{T^2(\tau,\bar{\tau})}\braket{O_1(v',\bar{v}')O_2(0)}_{T^2(\tau',\bar{\tau}')}=\mathrm{tr}(O_1(v,\bar{v})O_2(0)e^{-\beta H})\mathrm{tr}(O_1(v,'\bar{v}')O_2(0)e^{-\beta H})
\ee
Note that here the notion of late-time is different from section \ref{1ptlatetime}. In section \ref{1ptlatetime}, at late time the two tori on the boundary becomes large in time direction. Here two operators inserted on each torus becomes far away from each other. We take the twist of the Euclidean torus $s=0$ but we do not analytically continue $\beta$. Instead, we just analytically continue the location of the insertion $O_1$ on both boundaries 
\begin{align}
v&=i(\tfrac{\beta}{2}+it)=i\tfrac{\beta}{2}-t\quad\quad v'=i(\tfrac{\beta}{2}-it)=i\tfrac{\beta}{2}+t\\
\bar{v}&=-i(\tfrac{\beta}{2}+it)=-i\tfrac{\beta}{2}+t\quad\quad\bar{v}'=-i(\tfrac{\beta}{2}-it)=-i\tfrac{\beta}{2}-t
\end{align}
plugging these into (\ref{2ptresult}) we get
\begin{multline}
\overline{\braket{O_1(v,\bar{v})O_2(0)}_{T^2(\tau,\bar{\tau})}\braket{O_1(v',\bar{v}')O_2(0)}_{T^2(\tau',\bar{\tau}')}}\\
=2\braket{O_1(i\tfrac{\beta}{2}-t,-i\tfrac{\beta}{2}-t)O_2(0)}^L_{T^2(i\beta,-i\beta)}\braket{O_1(i\tfrac{\beta}{2}+t,-i\tfrac{\beta}{2}+t)O_2(0)}^L_{T^2(i\beta,-i\beta)}
\end{multline}
Observe that one torus has one insertion at $0$ and another insertion at $i\frac{\beta}{2}-\lfloor t\rfloor$. The other torus has one insertion at $0$ and another insertion at $i\frac{\beta}{2}+\lfloor t\rfloor$. Here $\lfloor\cdot\rfloor$ means the fractional part of $t$. Thus, as $t$ becomes large, the arguments don't change much, so $\overline{\braket{O_1(v,\bar{v})O_2(0)}_{T^2(\tau,\bar{\tau})}\braket{O_1(v',\bar{v}')O_2(0)}_{T^2(\tau',\bar{\tau}')}}$ does not decay at late time.

\subsection{Single-boundary quotients of a torus wormhole}
\label{3dcrosscap}
So far we have discussed two-boundary configurations with large Lorentzian separation that do not decay with time. In this section we calculate $\Z_2$ quotients of a torus wormhole from both the bulk side (section \ref{crosscapgravity}) and the boundary side (section \ref{crosscapcft}). These configurations each has one boundary which is a torus, so they contribute to the torus two-point function. And then using the result from \ref{2ptlatetime}, we observe that these contributions do not decay at late time. 

\subsubsection{Semi-classical gravity calculation}
\label{crosscapgravity}

\begin{figure}[H]
\centering
\includegraphics[width=0.25\textwidth]{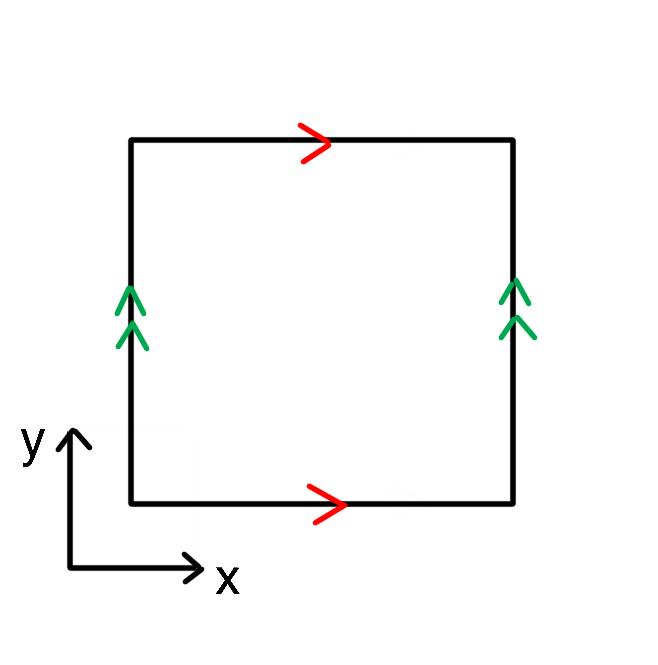}

$(x,y)\sim(x+\Z,y+\Z)$
\caption{A torus represented as a square with sides identified.}
\label{torussquare}
\end{figure}

Let us classify ways of doing $\Z_2$ quotient of a torus wormhole that gives a smooth geometry with one torus boundary. In other words, we classify $\Z_2$ symmetries of the torus wormhole such that the two boundaries are mapped to each other. Our trick of doing the classification is to focus on the zero-curvature slice in the middle of the torus wormhole. This zero-curvature slice is a torus that is mapped to itself under the $\Z_2$ symmetry. We represent this torus as a square with sides identified as shown in figure \ref{torussquare}, i.e. $(x,y)\sim(x+\Z,y+\Z)$.

There are two inequivalent $\Z_2$ symmetries and doing a quotient results in either a torus $T^2$ or a Klein bottle $K^2$ as shown in table \ref{quotienttable}. Without loss of generality, the fundamental region is the part to the left of the blue dashed line.

\begin{table}[H]
\centering
\begin{tabular}{c | c}
$T^2$&$K^2$\\
\hline
\\
\includegraphics[width=0.25\textwidth]{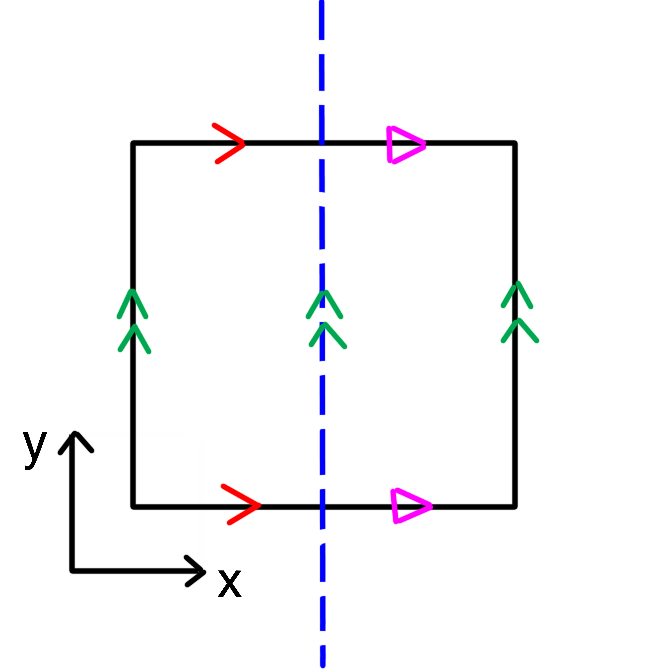} & \includegraphics[width=0.25\textwidth]{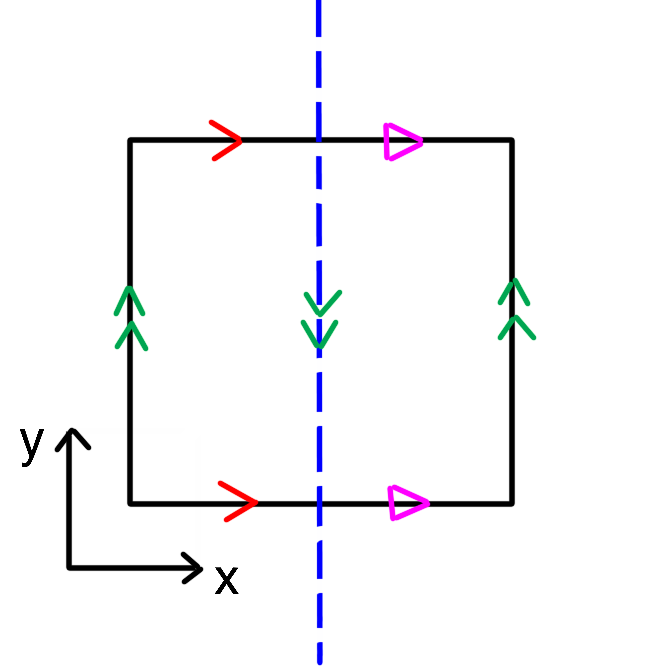}\\
\\
$(x,y)\sim(x+1/2,y)$&$(x,y)\sim(x+1/2,-y)$
\end{tabular}
\caption{Ways of doing a $\Z_2$ quotient of a torus}
\label{quotienttable}
\end{table}

\begin{figure}[H]
\centering
\includegraphics[width=\textwidth]{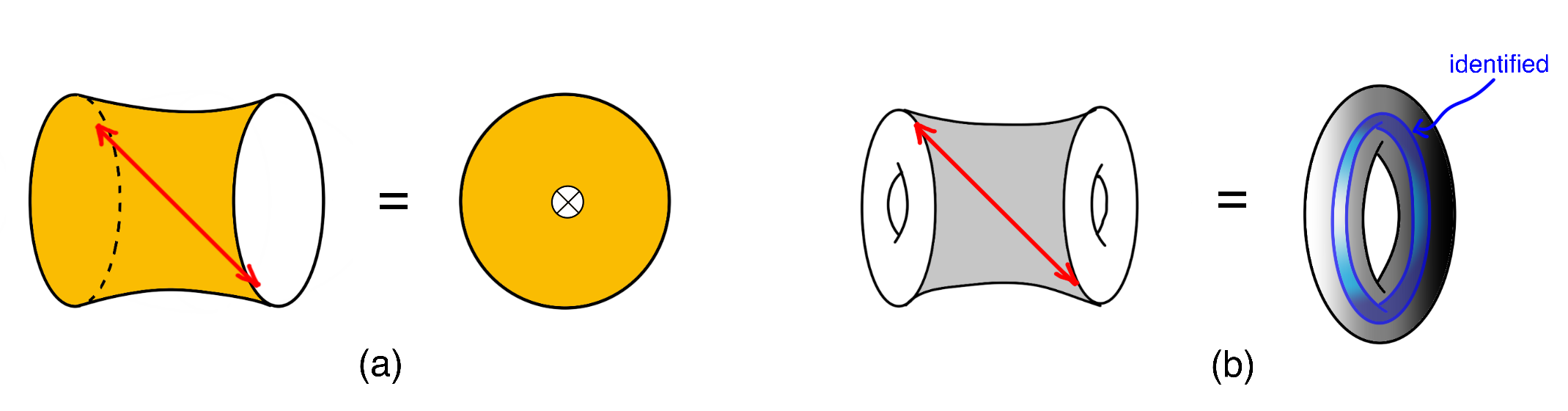}
\caption{(a) $\Z_2$ quotient of a cylinder in 2d (b) $\Z_2$ quotients of a torus wormhole in 3d}
\label{3dto2dcc}
\end{figure}

Going back to 3d, the above analysis tells us that there are two ways of doing $\Z_2$ quotient of a torus wormhole. Let us use the coordinate $(x,y,z)\in T^2\times [-1,1]$ for the torus wormhole. Thus these two $\Z_2$ quotient corresponds to (1) $(x,y,z)\sim(x+1/2,y,-z)$ and (2) $(x,y,z)\sim(x+1/2,-y,-z)$ respectively.\footnote{$(x,y,z)\sim(x+1/2,y,-z)$, $(x,y,z)\sim(x,y+1/2,-z)$, and $(x,y,z)\sim(x+1/2,y+1/2,-z)$ are equivalent identifications because we can change the ways we represent a torus on a $\mathbb{R}^2$ plane.} We should note that both these configurations contain a Mobius band $(x,0,z)\sim(x+1/2,0,-z)$. Now let us examine whether our two kinds of $\Z_2$ quotients are orientable or not. To do that we go back to the torus wormhole which is a 3d smooth manifold embedded in $\mathbb{R}^3$. We can think of the identifications as transition maps between two different elements in the atlas of the torus wormhole. The $\Z_2$ quotient is orientable iff the atlas is oriented. The atlas is oriented (unoriented) if the Jacobi determinant is positive (negative). 
\begin{enumerate}[(1)]
\item $(x,y,z)\sim(x+1/2,y,-z)$. This has Jacobi det $=-1$, which means it's a non-orientable 3d geometry. In particular, this configuration is equivalent to a Mobius strip $\times\,S^1$.
\item $(x,y,z)\sim(x+1/2,-y,-z)$. This has Jacobi det $=1$, which means it's an orientable 3d geometry. Intuitively, it has a twist in the $z$ direction and another twist in the $y$-direction, they cancel each other. 
\end{enumerate}
Notice that there is an analogy between $\Z_2$ quotients of a 3d torus wormhole and the $\Z_2$ quotient of a 2d cylinder as shown in figure \ref{3dto2dcc}. $\Z_2$ quotient of a 2d cylinder is the same as a disk with a crosscap inserted as shown in figure \ref{3dto2dcc}(a). $\Z_2$ quotients of a 3d torus wormhole are also equivalent to carving out a solid torus and then identify points on this torus as shown in figure \ref{3dto2dcc}(b). We should note that if we insert operators on the boundaries of torus wormhole and then take the $\Z_2$ quotient, the insertions need to be compatible with the corresponding $\Z_2$ symmetry.

\begin{figure}[H]
\centering
\includegraphics[width=0.3\textwidth]{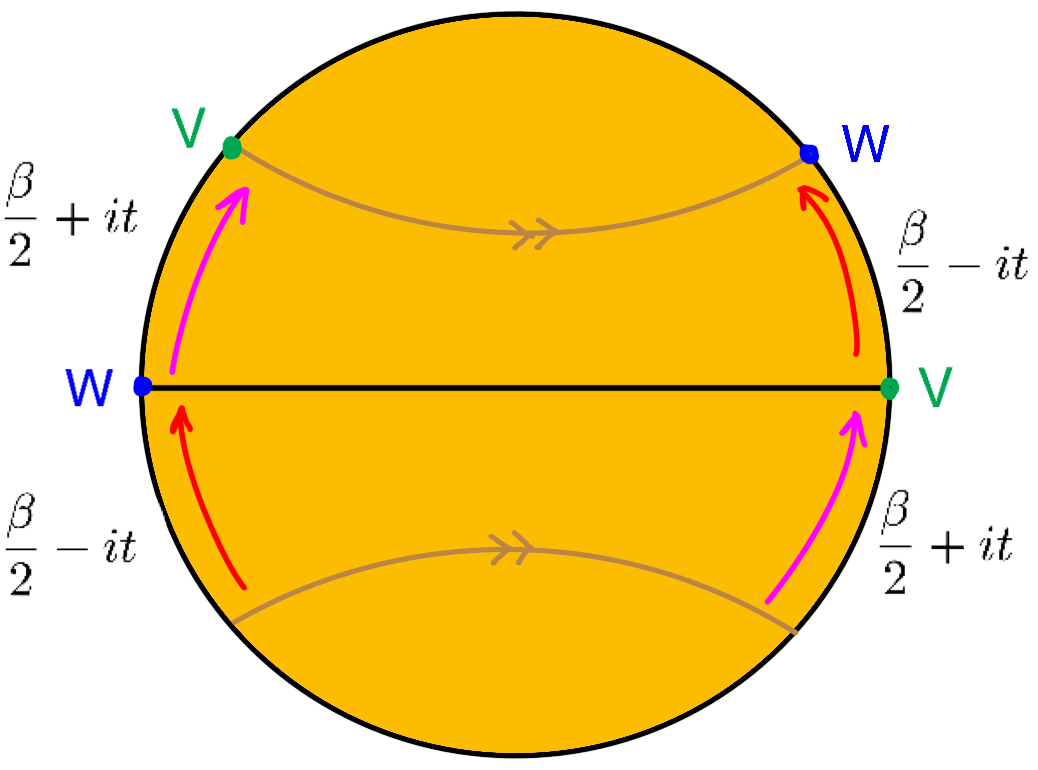}
\caption{In 2d, we get a cylinder from the hyperbolic disk by identifying the two brown geodesics. Then we insert two pairs of operators $V$ and $W$ on the boundary.}
\label{2dcylinder}
\end{figure}

Before analyzing the insertions in 3d torus wormhole, let us recall a similar situation in 2d where we insert four operators on the cylinder that respect the antipodal map so that we can do the antipodal identification for the cylinder later. The configuration of insertions compatible with the $\Z_2$ symmetry is shown in figure~\ref{2dcylinder}. The two $V$'s are antipodal of each other and the two $W$'s are also antipodal of each other. We analytically continue the Euclidean distance between $V$ and $W$ to $\frac{\beta}{2}+it$ and $\frac{\beta}{2}-it$. In particular, we can check that the configuration shown in figure~\ref{2dcylinder} makes sense by observing that the left boundary and right boundary of the cylinder gives the same two-point correlator.
\begin{align}
\mathrm{LHS}&=\mathrm{tr}(e^{-(\frac{\beta}{2}-it)H}Ve^{-(\frac{\beta}{2}+it)H}W)\\
\mathrm{RHS}&=\mathrm{tr}(e^{-(\frac{\beta}{2}+it)H}We^{-(\frac{\beta}{2}-it)H}V)
\end{align}
so LHS=RHS.

\noindent If we use the $(x,z)$ coordinate to denote $S^1\times[-1,1]$, the insertions are at
\begin{align}
(x_V,z_V)&=(0,-1)  &&(x'_V,z'_V)=(\frac{\beta}{2},1)\\
(x_W,z_W)&=(\frac{\beta}{2}+it,-1)  &&(x'_W,z'_W)=(\beta+it,1)
\end{align}
These satisfy $\Z_2$ symmetry because $(x'_v,z'_v)=(x_v+\beta/2,-z_v)$ and $(x'_w,z'_w)=(x_w+\beta/2,-z_w)$.

Analogously in 3d we again insert four operators so that they satisfy the $\Z_2$ symmetries, i.e. $(x'_v,y'_v,z'_v)=(x_v+\beta/2,\pm y_v,-z_v)$ and $(x'_w,y'_w,z'_w)=(x_w+\beta/2,\pm y_w,-z_w)$
\begin{align}
(x_V,y_V,z_V)&=(0,0,-1)  &&(x'_V,y'_V,z'_V)=(\frac{\beta}{2},0,1)\\
(x_W,y_W,z_W)&=(\frac{\beta}{2}+it,0,-1)  &&(x'_W,y'_W,z'_W)=(\beta+it,0,1)
\end{align}
Note that both (1) and (2) $\Z_2$ quotients are compatible with these operator insertions because the $y$-coordinates of all four insertions are zero and the only difference between (1) and (2) is from the $y$-coordinates. Again the distance between $V$ and $W$ are $\frac{\beta}{2}+it$ and $\frac{\beta}{2}-it$. If the tori are given by $T^2(\tau,\bar{\tau})=T^2(i\beta,-i\beta)$ and $T^2(\tau',\bar{\tau}')=T^2(i\beta,-i\beta)$ respectively. The insertions should be $V$ at $0$ and $W$ at $(v,\bar{v})=\big(i(\frac{\beta}{2}+it),-i(\frac{\beta}{2}+it)\big)$ for the left torus, and should be $W$ at $0$ and $V$ at $(v',\bar{v}')=\big(i(\frac{\beta}{2}-it),-i(\frac{\beta}{2}-it)\big)$ for the right torus. 

\begin{figure}[H]
\centering
\includegraphics[width=0.5\textwidth]{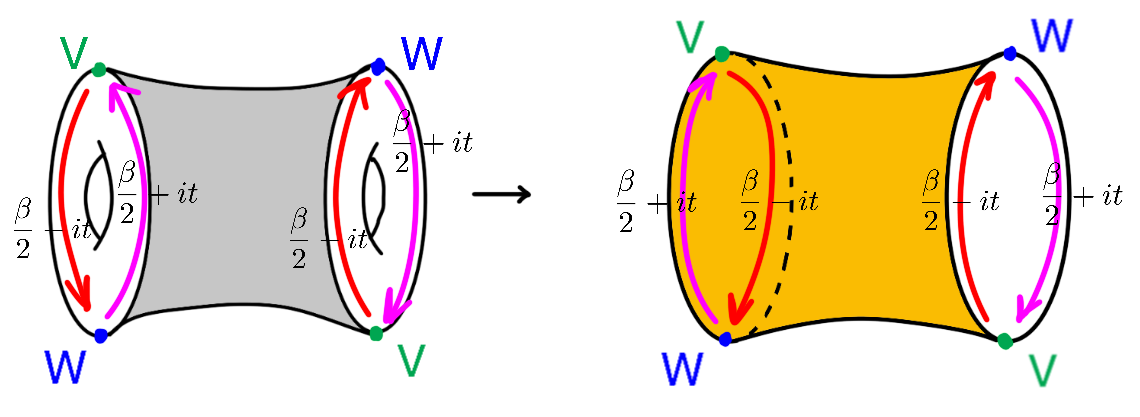}
\caption{We insert operators the same way for a 3d torus wormhole as for a 2d cylinder.}
\label{3dto2dop}
\end{figure}

There is one more subtlety to the configuration. For the operators to satisfy the $\Z_2$ symmetries, we should actually insert $TVT^{-1}$ and $TWT^{-1}$ for (1) and $(RT)V(RT)^{-1}$ and $(RT)W(RT)^{-1}$ for (2) where $T$ denotes time-reversal symmetry and $R$ denotes reflection. 

Thus the torus wormhole contribution to the averaged product of two torus two-point functions we are calculating is 
\begin{enumerate}[(1)]
\item 
\be
\overline{\braket{W(v,\bar{v})V(0)}_{T^2(\tau,\bar{\tau})}\braket{TV(v',\bar{v}')W(0)T^{-1}}_{T^2(\tau',\bar{\tau}')}}\label{product1}
\ee
\item
\be
\overline{\braket{W(v,\bar{v})V(0)}_{T^2(\tau,\bar{\tau})}\braket{(RT)V(v',\bar{v}')W(0)(RT)^{-1}}_{T^2(\tau',\bar{\tau}')}}\label{product2}
\ee
\end{enumerate}
Now we take $V=O_1$ and discuss the two cases separately
\begin{enumerate}[(1)]
\item (\ref{product1}) is only nonzero when 
\be
W=TO_1T^{-1}
\ee
Using (\ref{2ptresult}), we get
\begin{multline}
\overline{\braket{TO_1(v,\bar{v})T^{-1}O_1(0)}_{T^2(\tau,\bar{\tau})}\braket{TO_1(v',\bar{v}')T^{-1}O_1(0)}_{T^2(\tau',\bar{\tau}')}}\\
=2\braket{TO_1(i\tfrac{\beta}{2}-t,-i\tfrac{\beta}{2}-t)T^{-1}O_1(0)}^L_{T^2(i\beta,-i\beta)}\braket{TO_1(i\tfrac{\beta}{2}+t,-i\tfrac{\beta}{2}+t)T^{-1}O_1(0)}^L_{T^2(i\beta,-i\beta)}
\end{multline}
This is given by two quotient wormhole contributions glued together and semiclassically we can write
\be\label{lkjsdlfkj}
e^{-S_{\text{wormhole}}}=e^{-2S_{\text{quotient}}}
\ee
Thus quotient (1) contributes to the torus two-point function by
\begin{align}
&\overline{\braket{TO_1T^{-1}O_1}}_{(1)}=\includegraphics[valign=c,width=0.15\textwidth]{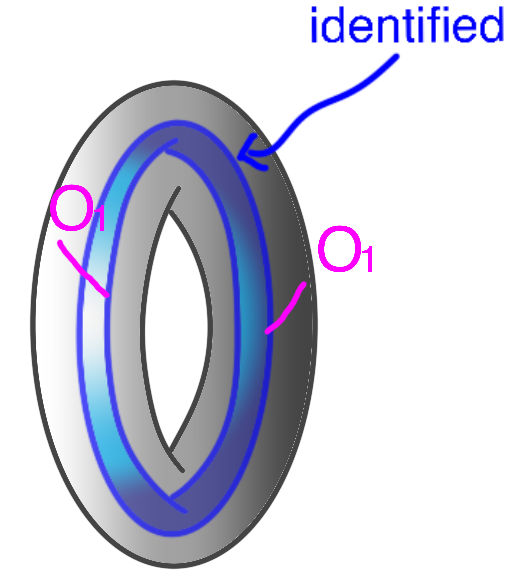}\\
&\approx_{\text{classical}}\sqrt{\overline{\braket{TO_1(v,\bar{v})T^{-1}O_1(0)}_{T^2(\tau,\bar{\tau})}\braket{TO_1(v',\bar{v}')T^{-1}O_1(0)}_{T^2(\tau',\bar{\tau}')}}}\\
&=\sqrt{2\braket{TO_1(i\tfrac{\beta}{2}-t,-i\tfrac{\beta}{2}-t)T^{-1}O_1(0)}_{T^2(i\beta,-i\beta)}\braket{TO_1(i\tfrac{\beta}{2}+t,-i\tfrac{\beta}{2}+t)T^{-1}O_1(0)}_{T^2(i\beta,-i\beta)}}\label{semiclassicalleq}
\end{align}
and this is non-decaying over time. We should note that this equation is only valid in the leading (zero loop) classical approximation, because we are just using the relationship between the classical actions (\ref{lkjsdlfkj}). Also since
\begin{multline}
\overline{\braket{O_1(v,\bar{v})O_2(0)}_{T^2(\tau,\bar{\tau})}\braket{O_1(v',\bar{v}')O_2(0)}_{T^2(\tau',\bar{\tau}')}}=\\
2\big|\int dh_p dh_q\rho_0(h_p)\rho_0(h_q)C_0(h_1,h_p,h_q)C_0(h_2,h_p,h_q)\mathcal{F}_{11}^{g=1}(h_p,h_q;\tau,v)\mathcal{F}_{11}^{g=1}(h_p,h_q;\tau',v')\big|^2
\end{multline}
we can use saddle-point approximation (for a justification see Appendix \ref{saddle}) to get
\be
\overline{\braket{TO_1T^{-1}O_1}}_{(1)}\approx_{\text{classical}}|\int dh_p\rho_0(h_p)C_0(h_1,h_p,h_p)\mathcal{F}_{11}^{g=1}(h_p,h_p;\tau,v)|^2\label{con1}
\ee

\item
(\ref{product2}) is only nonzero when 
\be
W=(RT)O_1(RT)^{-1}
\ee
Using (\ref{2ptresult}), we get
\begin{multline}
\overline{\braket{(RT)O_1(v,\bar{v})(RT)^{-1}O_1(0)}_{T^2(\tau,\bar{\tau})}\braket{(RT)O_1(v',\bar{v}')(RT)^{-1}O_1(0)}_{T^2(\tau',\bar{\tau}')}}\\
=2\braket{(RT)O_1(i\tfrac{\beta}{2}-t,-i\tfrac{\beta}{2}-t)(RT)^{-1}O_1(0)}^L_{T^2(i\beta,-i\beta)}\braket{(RT)O_1(i\tfrac{\beta}{2}+t,-i\tfrac{\beta}{2}+t)(RT)^{-1}O_1(0)}^L_{T^2(i\beta,-i\beta)}
\end{multline}
Quotient (2) contributes to the torus two-point function by
\begin{align}
&\hspace{-1cm}\overline{\braket{(RT)O_1(RT)^{-1}O_1}}_{(2)}=\includegraphics[valign=c,width=0.15\textwidth]{figures/torus2pt2.png}\\
&\hspace{-1cm}\approx_{\text{classical}}\sqrt{\overline{\braket{(RT)O_1(v,\bar{v})(RT)^{-1}O_1(0)}_{T^2(\tau,\bar{\tau})}\braket{(RT)O_1(v',\bar{v}')(RT)^{-1}O_1(0)}_{T^2(\tau',\bar{\tau}')}}}\\
&\hspace{-1cm}=\sqrt{2\braket{(RT)O_1(i\tfrac{\beta}{2}-t,-i\tfrac{\beta}{2}-t)(RT)^{-1}O_1(0)}_{T^2(i\beta,-i\beta)}\braket{(RT)O_1(i\tfrac{\beta}{2}+t,-i\tfrac{\beta}{2}+t)(RT)^{-1}O_1(0)}_{T^2(i\beta,-i\beta)}}\label{semiclassicalleq}
\end{align}
and this is non-decaying over time. Again this is only valid in the leading (zero loop) classical approximation. And again using saddle-point approximation
\be
\overline{\braket{(RT)O_1(RT)^{-1}O_1}}_{(2)}\approx_{\text{classical}}|\int dh_p\rho_0(h_p)C_0(h_1,h_p,h_p)\mathcal{F}_{11}^{g=1}(h_p,h_p;\tau,v)|^2
\ee
But $RT$ acting on $O_1$ is equivalent to a $180^\circ$ rotation so
\begin{align}
\overline{\braket{O_1O_1}}_{(2)}&=(-1)^{s_1}\overline{\braket{(RT)O_1(RT)^{-1}O_1}}_{(2)}\\
&\approx_{\text{classical}}(-1)^{s_1}|\int dh_p\rho_0(h_p)C_0(h_1,h_p,h_p)\mathcal{F}_{11}^{g=1}(h_p,h_p;\tau,v)|^2\label{con2}
\end{align}
\end{enumerate}
(\ref{con1}) and (\ref{con2}) are results of a classical approximation. However, it's interesting to contemplate that whether they are actually exact equations.

\subsubsection{CFT calculation}
\label{crosscapcft}

Before calculating the $\Z_2$ quotients of a torus wormhole from the boundary side, let us examine the boundary ensemble in \cite{Chandra:2022bqq} more closely. 
\be
\overline{c_{abc}c^*_{def}}=C_0(h_a,h_b,h_c)C_0(\bar{h}_a,\bar{h}_b,\bar{h}_c)(\delta_{ad}\delta_{be}\delta_{cf}+(-1)^{s_a+s_b+s_c}\delta_{ad}\delta_{bf}\delta_{ce}+\text{4 more terms})\label{Hartmanens}
\ee
where the remaining four terms are cyclic permutations of the first two terms. Let's start by writing the OPE coefficients of two heavy operators $O_i, O_j$ and one light operator $O_\alpha$ in operator form
\be
c_{i\alpha j}=\braket{i|O_\alpha|j}=\braket{O_iO_\alpha O_j}
\ee
Here all operators that are Hermitian. We should note that 
\be
c_{i\alpha j}=\braket{i|O_\alpha|j}=\braket{i|O_\alpha^\dagger|j}=\braket{j|O_\alpha|i}^*=c_{j\alpha i}^*
\ee
We now show that an OPE coefficient $c_{123}$ is real if $s_1+s_2+s_3$ is even and purely imaginary if $s_1+s_2+s_3$ is odd. If the operator $O$ is a symmetric traceless tensor of spin $s=h-\bar{h}$, we can write it in component form as
\be
O(x,J)=O_{\mu_1,\cdots,\mu_s}J^{\mu_1}\cdots J^{\mu_s}
\ee
Let $\mathrm{Rot}$ denote rotation by $180^\circ$, define $\ket{1}=O_1(x_1,J_1)\ket{0}$ and $\ket{3}=O_3(x_3,J_3)\ket{0}$, then three-point function is given by
\begin{multline}
\braket{1|O_2(x_2,J_2)|3}=\mathrm{Rot}\braket{1|O_2(x_2,J_2)|3}\\=(-1)^{s_1+s_2+s_3}\braket{3|O_2(x_2,J_2)|1}=(-1)^{s_1+s_2+s_3}\braket{1|O_2(x_2,J_2)|3}^*
\end{multline}
where in the second line the prefactor comes from rotating the spin while the braket comes from rotating operators keeping spin fixed. Thus
\be
c_{i\alpha j}^*=(-1)^{s_i+s_\alpha+s_j}c_{i\alpha j}
\ee
The ensemble of OPE coefficients can be justified by considering a simple model of Haar random matrices. We introduce a change of basis by unitary matrix $u$ and write 
\be
c_{i\alpha j}=(u^\dagger O_\alpha u)_{ij}
\ee
where $u$ is a unitary matrix and they form a Gaussian unitary ensemble (GUE) \cite{Dyson}. We can justify writing the change of basis $u$ by thinking that there are two sets of basis: local basis and energy eigenbasis. In local basis, the operator $O_\alpha$ is simple but in the energy eigenbasis $O_\alpha$ need to be transformed by some rather complicated change of basis, so we can take those complicated matrix $u$ to be random. We can then model averaging over ensemble of OPE coefficients as integral over random unitary matrices. Recall we have the leading order contribution to Weingarten's formula \cite{Weingartens}
\begin{equation}
\int du\,u_{i_1}^{\phantom{i_1}j_1}u_{i_2}^{\phantom{i_2}j_2}(u^\dagger)_{k_1}^{\phantom{k_1}l_1}(u^\dagger)_{k_2}^{\phantom{k_2}l_2}\approx\frac{1}{L^2}\big(\delta_{i_1}^{l_1}\delta_{i_2}^{l_2}\delta_{k_1}^{j_1}\delta_{k_2}^{j_2}+\delta_{i_1}^{l_2}\delta_{i_2}^{l_1}\delta_{k_1}^{j_2}\delta_{k_2}^{j_1}\big)\label{wformula}
\end{equation}
Now let us review how to derive this formula. The general idea is we observe that $u$ is a projector onto invariants. Thus we can first find the invariant states then we can just write the integral formula immediately. More specifically, denote our Hilbert space by $\H$, then our random matrices as maps $u,u^\dagger:\H\rightarrow\H$ and $u^*,u^T:\H^*\rightarrow\H^*$. 
\be
\ket{i}\mapsto u_i^{\phantom{i}j}\ket{j}\quad\quad \bra{i}\mapsto (u^*)^i_{\phantom{i}j}\bra{j}
\ee
Let us denote elements of $\{\H, \H, \H^*,\H^*\}$ using the indices $1,2,3,4$ respectively. Then the following state is invariant under evolutions by Haar random matrices\footnote{The states $\ket{i}$ are orthogonal in the leading order approximation of Weingarten's formula.}
\be
\ket{13}=\sum_{i}\ket{i}\bra{i}
\ee
because
\be
u\otimes u^*\ket{13}=\sum_i(u\ket{i})(u^*\bra{i})=\sum_{i,j,k}\ket{j}\braket{j|u|i}\braket{i|u^\dagger|k}\bra{k}=\sum_{j,k}\ket{j}\braket{j|uu^\dagger|k}\bra{k}=\sum_j\ket{j}\bra{j}=\ket{13}
\ee
Note in particular that $\ket{13}$ is the identity operator which is basis-independent, and so well-defined. The same hold for $\ket{24}, \ket{14}, \ket{23}$ so
\begin{align}
\int du\,u_{i_1}^{\phantom{i_1}j_1}u_{i_2}^{\phantom{i_2}j_2}(u^\dagger)_{k_1}^{\phantom{k_1}l_1}(u^\dagger)_{k_2}^{\phantom{k_2}l_2}&=\int du\,u_{i_1}^{\phantom{i_1}j_1}u_{i_2}^{\phantom{i_2}j_2}(u^*)_{\phantom{l_1}k_1}^{l_1}(u^*)_{\phantom{l_2}k_2}^{l_2}\\
&\approx\frac{1}{L^2}\big(\delta_{i_1}^{l_1}\delta_{i_2}^{l_2}\delta_{k_1}^{j_1}\delta_{k_2}^{j_2}+\delta_{i_1}^{l_2}\delta_{i_2}^{l_1}\delta_{k_1}^{j_2}\delta_{k_2}^{j_1}\big)
\end{align}

For two boundary dimensions, we should modify (\ref{wformula}) by adding reflection + time-reversal, i.e. RT symmetry.\footnote{Relativistic theories always have this symmetry. It's sometimes called CPT symmetry.} We want to know how the ensemble of OPE coefficients would change if we add RT symmetry (i.e. $RT$ commuting with $u$ or $u^\dagger RT u=RT$) to GUE. Recall that Lorentzian time-reversal $T$ is antilinear and antiunitary. Since we know that complex conjugate $K$ is antilinear and antiunitary \footnote{Complex conjugate operator is an antilinear and antiunitary operator $K:\mathcal{H}\rightarrow\mathcal{H}$ because $K\sum\alpha_i\ket{i}=\sum\alpha_i^*\ket{i}$ and if we write $\ket{\psi}=\sum\alpha_i\ket{i}, \quad \ket{\chi}=\sum\beta_i\ket{i}$ then we have the inner product $\braket{K\chi| K\psi}=\sum_{i,j}(\bra{j}\beta_j)(\alpha^*_i\ket{i})=\sum_i\alpha_i^*\beta_i\braket{i|i}=\sum_{i,j}(\bra{j}\alpha^*_j)(\beta_i\ket{i})=\braket{\psi|\chi}$.}, it is natural to model $T$ with a factor of $K$ in it. With $RT$ symmetry, there are more invariant states
\be
\ket{12}'=\sum_{i}\ket{i}RTK\ket{i}
\ee
since 
\be
u\otimes u\ket{12}'=\sum_i\ket{i}uRTKu^T\ket{i}=\sum_i\ket{i}uRTu^\dagger K\ket{i}=\sum_i\ket{i}RTK\ket{i}=\ket{12}'
\ee
Therefore, time-reversal symmetry adds a term to the Weingarten's formula
\begin{multline}
\int du\,u_{i_1}^{\phantom{i_1}j_1}u_{i_2}^{\phantom{i_2}j_2}(u^\dagger)_{k_1}^{\phantom{k_1}l_1}(u^\dagger)_{k_2}^{\phantom{k_2}l_2}\approx\frac{1}{L^2}\big(\delta_{i_1}^{l_1}\delta_{i_2}^{l_2}\delta_{k_1}^{j_1}\delta_{k_2}^{j_2}+\delta_{i_1}^{l_2}\delta_{i_2}^{l_1}\delta_{k_1}^{j_2}\delta_{k_2}^{j_1}\\+(RTK)^{-1}_{i_1,i_2}(RTK)^{j_1,j_2}(RTK)^{-1}_{k_1,k_2}(RTK)^{l_1,l_2}\big)
\end{multline}
Thus the averaged product of two OPE coefficients with one light and two heavy operators is
\begin{align}
\overline{c_{i\alpha j}c_{k\beta l}^*}&=\int du(u^\dagger O_\alpha u)_{ij}(u^\dagger O_\beta u)_{lk}\\
&=\frac{1}{L^2}\left(\tr(O_\alpha O_\beta )\delta_{ik}\delta_{jl}+\mathrm{Tr}\big((RTK)O^T_\alpha (RTK)^{-1} O_\beta\big)(RTK)^{-1}_{il}(RTK)_{jk}\right)\\
&=\frac{1}{L^2}\left(\tr(O_\alpha O_\beta )\delta_{ik}\delta_{jl}+\mathrm{Tr}\big((RT)O^\dagger_\alpha (RT)^{-1} O_\beta\big)(RTK)^{-1}_{il}(RTK)_{jk}\right)\\
&=\frac{1}{L}\left(\delta_{\alpha\beta}\delta_{ik}\delta_{jl}+(-1)^{s_i+s_\alpha+s_j}\delta_{\alpha\beta}\delta_{il}\delta_{jk}\right)\label{guecft}
\end{align}
where in the last line we used two facts: First, $RT$ acting on $O_\alpha$ by conjugation is equivalent to rotation by $180^\circ$ so
\be
\mathrm{Tr}\big((RT)O^\dagger_\alpha (RT)^{-1} O_\beta\big)=(-1)^{s_\alpha}\mathrm{Tr}(O_\alpha^\dagger O_\beta)=(-1)^{s_\alpha}\delta_{\alpha\beta}
\ee
Second, we can write $(RTK)^{-1}_{il}$ and $(RTK)_{jk}$ in braket notation as
\begin{align}
(RTK)^{-1}_{il}&=\left(\sum_n\bra{n}(RTK)^{-1}\bra{n}\right)\ket{i}\ket{l}=\sum_n\braket{i|(RT)^{-1}|n}\braket{n|l}=\braket{0|O_i^\dagger (RT)^{-1}O_l|0}=(-1)^{s_i}\delta_{il}\\
(RTK)_{jk}&=\bra{j}\bra{k}\left(\sum_n\ket{n}RTK\ket{n}\right)=\sum_n\braket{j|n}\braket{n|RT|k}=\braket{0|O^\dagger_jRTO_k|0}=(-1)^{s_j}\delta_{jk}
\end{align}
Thus we have reproduced (\ref{Hartmanens}).

Now we want to know how the ensemble of OPE coefficients would change if we add time-reversal symmetry without reflection $R$ (i.e. the time-reversal operator $T$ commuting with $u$ i.e. $u^\dagger T u=T$) to GUE. With $T$ symmetry, we have invariant state
\be
\ket{12}=\sum_{i}\ket{i}TK\ket{i}
\ee
because
\be
u\otimes u\ket{12}=\sum_i\ket{i}uTKu^T\ket{i}=\sum_i\ket{i}uTu^\dagger K\ket{i}=\sum_i\ket{i}TK\ket{i}=\ket{12}
\ee
Therefore, time-reversal symmetry adds a term to the Weingarten's formula
\be
\int du\,u_{i_1}^{\phantom{i_1}j_1}u_{i_2}^{\phantom{i_2}j_2}(u^\dagger)_{k_1}^{\phantom{k_1}l_1}(u^\dagger)_{k_2}^{\phantom{k_2}l_2}\supset(TK)^{-1}_{i_1,i_2}(TK)^{j_1,j_2}(TK)^{-1}_{k_1,k_2}(TK)^{l_1,l_2}
\ee
This would add an additional term to the averaged product of OPE coefficients 
\begin{align}
\overline{c_{i\alpha j}c_{k\beta l}^*}&\supset \frac{1}{L^2}\mathrm{Tr}\big((TK)O^T_\alpha (TK)^{-1} O_\beta\big)(TK)^{-1}_{il}(TK)_{jk}\\
&=\frac{1}{L^2}\mathrm{Tr}(TO_\alpha^\dagger T^{-1}O_\beta)(TK)^{-1}_{il}(TK)_{jk}\\
&=\begin{cases}\frac{1}{L}(TK)^{-1}_{il}(TK)_{jk}&O_\alpha=TO_\beta T^{-1}\\0&\text{otherwise}\end{cases}
\end{align}
There are two kinds of anomalies $T^2=\pm1$. For $T^2=1$, we can just take $T=K$ and the condition $TuT^{-1}=u$ reduces $u^\dagger u=1$ to $u^Tu=1$, which is equivalent to saying $u$ is orthogonal, and the ensemble becomes GOE. Thus the added term to ensemble simplifies to
\be
\overline{c_{i\alpha j}c_{k\beta l}^*}\supset\begin{cases}\frac{1}{L}\delta_{il}\delta_{jk}&O_\alpha=TO_\beta T^{-1}\\0&\text{otherwise}\end{cases}\label{goecft}
\ee
For $T^2=-1$, we can take $T=K\omega$ where $\omega=\begin{pmatrix}0&1\\-1&0\end{pmatrix}$. The condition $TuT^{-1}=u$ reduces $u^\dagger u=1$ to $u^T\omega u=\omega$, which is equivalent to saying $u$ is symplectic, and the ensemble becomes GSE. Thus the added term to ensemble simplifies to
\be
\overline{c_{i\alpha j}c_{k\beta l}^*}\supset\begin{cases}\frac{1}{L}\omega^{-1}_{il}\omega_{jk}&O_\alpha=TO_\beta T^{-1}\\0&\text{otherwise}\end{cases}\label{gsecft}
\ee

Now we can use results from last subsection to calculate torus two-point functions. Note that on our 2d boundary, we always have $RT$ symmetry so the ensemble is given by (\ref{guecft}), which we now use to contract indices of the OPE coefficients
\begin{align}
\overline{\braket{O_1(v,\bar{v})O_1(0)}_{T^2(\tau,\bar{\tau})}}&=\sum_{p,q}\overline{c_{1pq}c_{1pq}^*}\left|\includegraphics[valign=c,width=0.12\textwidth]{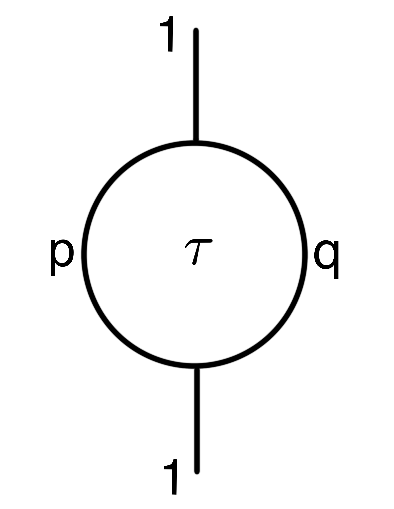}\right|^2=\sum_{p,q}\overline{c_{1pq}c_{1pq}^*}\big|\mathcal{F}_{11}^{(N)}(h_p,h_q;\tau,v)\big|^2\\
&=\sum_{p,q}\left(1+(-1)^{s_1+s_p+s_q}\delta_{pq}\right)\big|C_0(h_1,h_p,h_q)\big|^2\big|\mathcal{F}_{11}^{(N)}(h_p,h_q;\tau,v)\big|^2\\
&=\sum_{p,q}\big|C_0(h_1,h_p,h_q)\mathcal{F}_{11}^{(N)}(h_p,h_q;\tau,v)\big|^2\nonumber\\
&\quad\quad\quad+(-1)^{s_1}\sum_p\big|C_0(h_1,h_p,h_p)\mathcal{F}_{11}^{(N)}(h_p,h_p;\tau,v)\big|^2\\
&=\big|\int dh_p dh_q\rho_0(h_p)\rho_0(h_q)C_0(h_1,h_p,h_q)\mathcal{F}_{11}^{g=1}(h_p,h_q;\tau,v)\big|^2\nonumber\\
&\quad\quad\quad+(-1)^{s_1}\big|\int dh_p \rho_0(h_p)C_0(h_1,h_p,h_p)\mathcal{F}_{11}^{g=1}(h_p,h_p;\tau,v)\big|^2\label{cfccgue}
\end{align}
If we also add time-reversal symmetry, the ensemble becomes (\ref{goecft}). This would add another term to the two-point function
\begin{align}
\overline{\braket{TO_1(v,\bar{v})T^{-1}O_1(0)}_{T^2(\tau,\bar{\tau})}}&\supset \sum_p\overline{c_{1pp}^2}\left|\includegraphics[valign=c,width=0.12\textwidth]{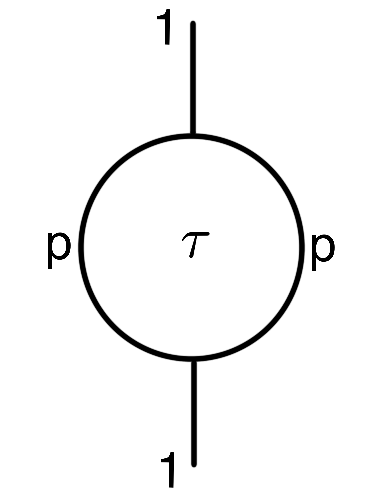}\right|^2=\sum_p\overline{c_{1pp}^2}|\mathcal{F}_{11}^{g=1}(h_p,h_p;\tau,v)|^2\\
&=|\int dh_p\rho_0(h_p)C_0(h_1,h_p,h_p)\mathcal{F}_{11}^{g=1}(h_p,h_p;\tau,v)|^2\label{cftcc}
\end{align}
Thus we see that the first term of (\ref{cfccgue}) is the original decaying result of \cite{Chandra:2022bqq}, the second term of (\ref{cfccgue}) matches contribution from (2) orientable $\Z_2$ quotient of torus wormhole on the gravity side (\ref{con2}), and (\ref{cftcc}) matches contribution from (1) non-orientable $\Z_2$ quotient (\ref{con1}).

\section{Comments on RT symmetry}
\label{discussion}

In this section, we first show that a generic relativistic quantum field theory with random matrix statistics should be of the GOE type for bosonic states and GSE for fermionic states, then we point out that the partition function of a torus wormhole calculated in \cite{Cotler:2020ugk, Cotler:2020hgz} needs another multiplicative factor of $2$.

Section \ref{crosscapcft} tells us that the CFT$_2$ ensemble proposed by \cite{Chandra:2022bqq} is inherently GOE for bosonic states and GSE for fermionic states since it contains $RT$ symmetry, and $RT$ symmetry is an anti-linear, anti-unitary symmetry that squares to $(-1)^F$ \cite{Witten:2015aba}. We should note that in 2d, $RT$ symmetry always exists (this can be understood as coming from the CPT theorem \footnote{C=charge conjugation, P=parity, T=time-reversal. In 2d, P=R and CT together is the time-reversal we are considering in RT.}). Inspired by the above observation, we claim that to the extent that a relativistic quantum field theory exhibits random matrix statistics it should be of the GOE type for bosonic states and of the GSE type for fermionic states. To start with, a relativistic quantum field theory exhibits random matrix statistics is of GUE, if the system has no additional symmetry, adding RT symmetry would make it into GOE or GSE. One caveat is that if the Hamiltonian has an additional symmetry which block diagonalize it, the RT symmetry could potentially exchange different blocks instead of acting on each individual block. We now show that the above situation does not happen. We can block diagonalize the Hamiltonian into different momentum blocks
\be
  \setlength{\arraycolsep}{0pt}
H=  \begin{pmatrix}
    \fbox{$H_{p_1}$} & 0 & 0  \\
    0 & \fbox{$H_{p_2}$} & 0 \\
    0 & 0 & \ddots \\
  \end{pmatrix}
\ee
In order to show the energy eigenvalue distribution is GOE or GSE, we need to make sure that each individual subblock of $H$ commutes with RT. This is equivalent to showing that momentum commutes with RT. Let $T_{\mu\nu}$ be the stress-energy tensor then the momentum is given by
\be
p(t)=\int dx\,T_{01}(x,t)
\ee
From here, we can show that
\be
RT\,p(0)=p(0)
\ee
Therefore, momentum commutes with CPT and we conclude that any chaotic CFT$_2$ has energy eigenvalue distribution a GOE for bosonic states and a GSE for fermionic states.

\begin{figure}[H]
\centering
\includegraphics[width=0.5\textwidth]{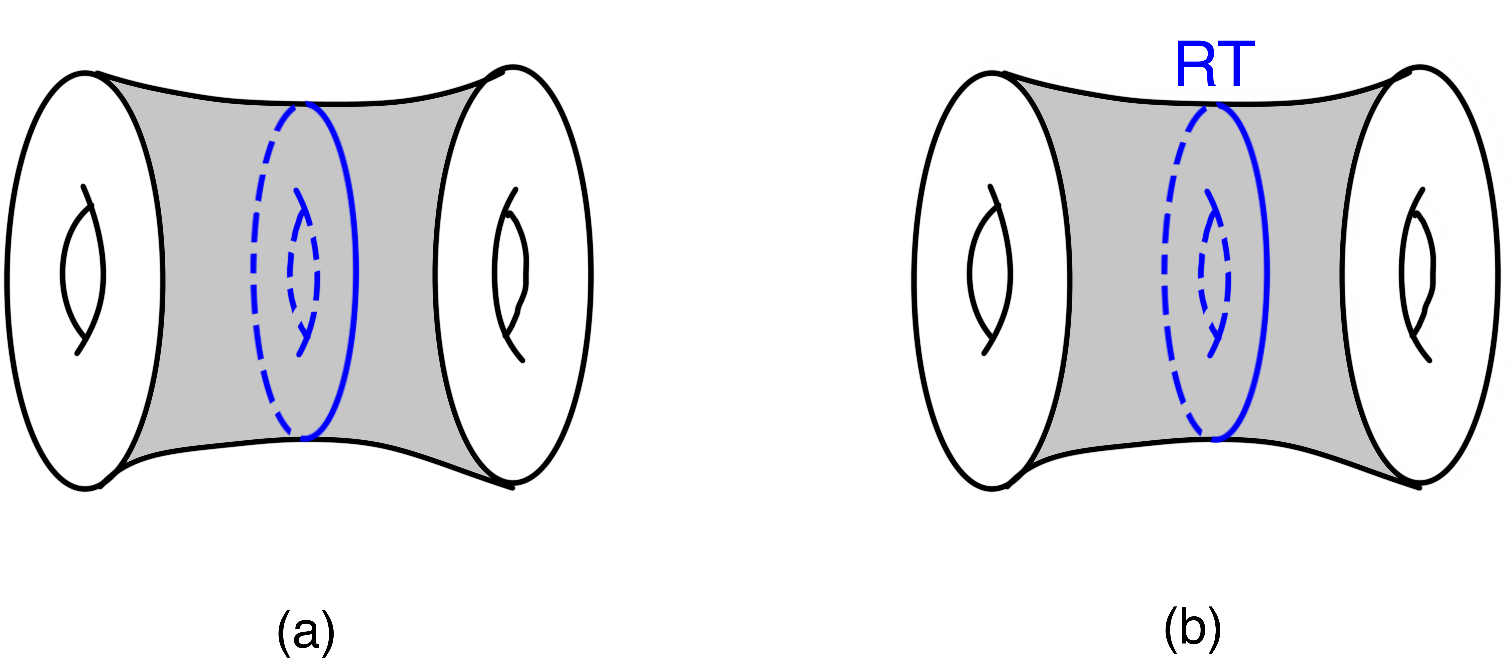}
\caption{(a) gluing two torus trumpets to get a torus wormhole (b) act on the right torus trumpet with RT and then glue to the left trumpet}
\label{torusRT}
\end{figure}

RT symmetry has implications for torus wormhole partition function studied in \cite{Cotler:2020ugk, Cotler:2020hgz}. In Euclidean AdS$_3$ bulk, having $RT$ symmetry means that in addition to the usual torus wormhole that can be obtained by gluing together two torus trumpets as shown in figure \ref{torusRT}(a), there should always exists a configuration that first act on the right torus trumpet with RT, i.e. rotate the right torus trumpet by $180^\circ$, and then glue it to the left torus trumpet as shown in figure \ref{torusRT}(b). This implies that the partition function of a torus wormhole calculated in \cite{Cotler:2020ugk, Cotler:2020hgz} needs another multiplicative factor of $2$.

\begin{figure}[H]
\centering
\includegraphics[width=0.6\textwidth]{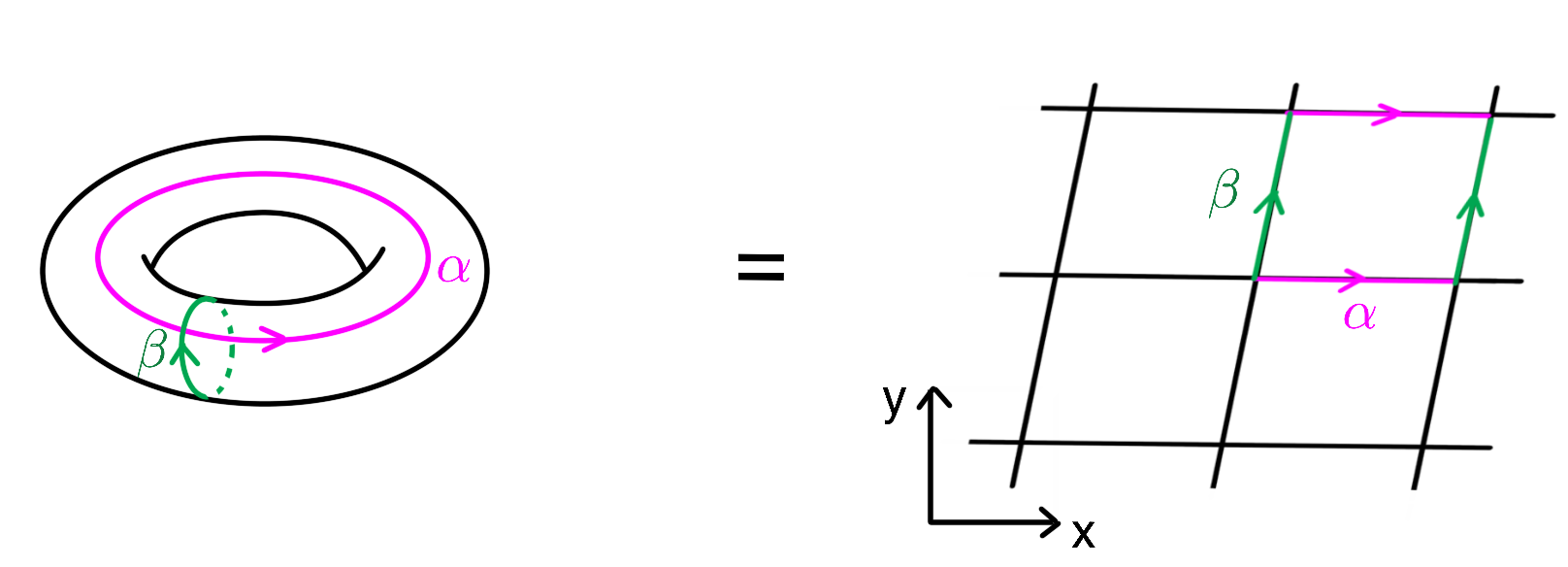}
\caption{$T^2$ is a quotient of the complex plane by a 2d lattice}
\label{torusMCG}
\end{figure}

Now we explain another way of understanding this which is through the Mapping Class Group. The Mapping Class Group (MCG) of a torus $T^2$ is given by the automorphisms of $T^2$, $\mathrm{Aut}(T^2)$, modding out by the path component of the identity in $\mathrm{Aut}(T^2)$, i.e.
\be
\mathrm{MCG}(T^2)=\mathrm{Aut}(T^2)/\mathrm{Aut}_0(T^2)
\ee
It turns out that 
\be
\mathrm{MCG}(T^2)=\mathrm{SL}(2,\Z)
\ee
Intuitively, this can be understood by focusing on how the two cycles of a torus $\alpha$ and $\beta$ get mapped. Recall that a torus is a quotient of the complex plane by a 2d lattice as shown in figure \ref{torusMCG}. 

\begin{figure}[H]
\centering
\includegraphics[width=0.8\textwidth]{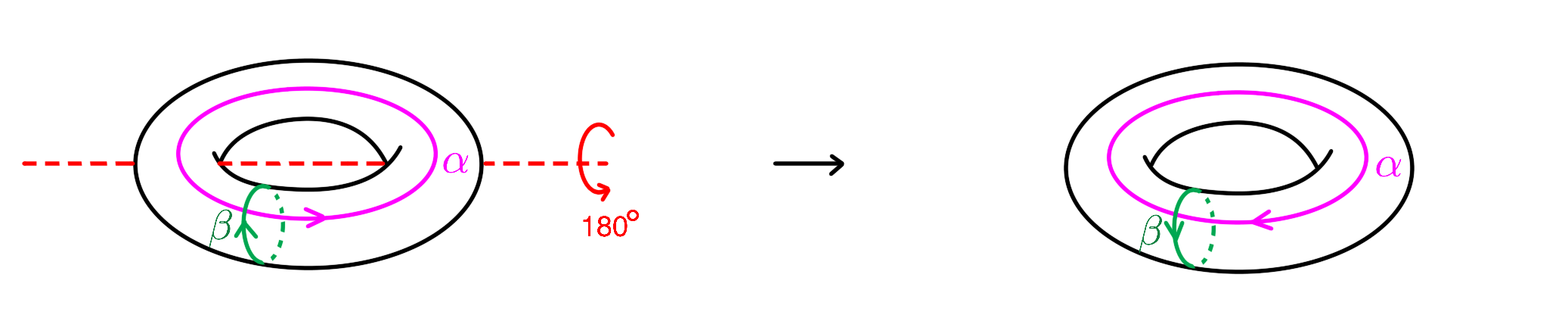}
\caption{A $180^\circ$ rotation flips the directions of both $\alpha$ and $\beta$ cycles of a torus.}
\label{torusreflection}
\end{figure}

\noindent The lattice has $\alpha$ and $\beta$ as basis vectors, and we think of them as complex numbers. We want to determine the number of ways this torus can be mapped to itself ignoring simply zooming in or out. $(\alpha,\beta)$ gets mapped to $(p\alpha, q\beta)$ where $p$ and $q$ are relatively prime integers with the same sign. We know that any such pair $(p,q)$ can be mapped from $(1,1)$ by a unique element of $\mathrm{SL}(2,\Z)$
\be
\begin{pmatrix}a&b\\c&d\end{pmatrix}\begin{pmatrix}1\\1\end{pmatrix}=\begin{pmatrix}p\\q\end{pmatrix}
\ee
Ignoring the sign of $(p,q)$. The complex structure of the torus is given by
\be
\tau=\frac{\alpha}{\beta}
\ee
and it get mapped by $\mathrm{PSL}(2,\Z)=\mathrm{SL}(2,\Z)/\{\pm1\}$ to
\be
\tau\mapsto\frac{a\tau+b}{c\tau+d}
\ee
However, we should note that this representation of $\mathrm{MCG}$ ignores the case where $(\alpha,\beta)$ gets mapped to $(-\alpha,-\beta)$. This is the case where the parallelogram formed by $(\alpha,\beta)$ get rotated by $180^\circ$. This $180^\circ$ rotation is not included in $\mathrm{PSL}(2,\Z)$ because it is identified with the identity element. However, with $RT$ symmetry, the directions of $\alpha$ and $\beta$ are important as shown in figure \ref{torusreflection}.

 Thus we should add in this element back, i.e. we should modify (\ref{CJsum}) to 
\be
Z_{T^2\times I}(\tau_1,\bar{\tau}_1,\tau_2,\bar{\tau}_2)=\sum_{\gamma\in\mathrm{SL}(2;\Z)}\tilde{Z}(\tau_1,\bar{\tau}_1,\gamma\tau_2,\gamma\bar{\tau}_2)
\ee
This gives an additional multiplicative factor of two in the result. In large spin, this result would again reduce to a double-trumpet in JT but with an additional factor of two, which is consistent with JT with time-reversal symmetry added\footnote{Time-reversal symmetry adds the double-trumpet glued from two trumpets but with a reflection on one of the interfaces being glued.} \cite{Stanford:2019vob}. Also note that this is more consistent with (\ref{1ptCJ}), if we take $s_1$ even and large.

\section*{Acknowledgements} 
I want to give special thanks to Douglas Stanford for patient guidance, extensive discussions, and inspiring comments throughout this project. I am also grateful to Eleny Ionel, Steven Kerckhoff, Raghu Mahajan, Henry Maxfield, and Xiaoliang Qi for discussions. 

\appendix

\section{JT gravity}
\label{2dappendix}

Calculations in the main text parallel those in JT gravity. In this Appendix, we give analogous results in JT gravity following the order of presentation of the main text. For a more detailed presentation see \cite{Yan:2022nod}.

In 2d, correlators are studied in the context of Jackiw-Teitelboim (JT) gravity \cite{Teitelboim, Jackiw, AlmheiriPolchinski} which is dual to an ensemble of quantum mechanical systems on the boundary \cite{Saad:2019lba, Stanford:2019vob} and can be described by Random Matrix Theory (RMT). Saad \cite{Saadsingleauthor} \footnote{continuing the idea of \cite{Blommaert:2019hjr}} computed bosonic two-point correlation functions using the techniques developed by Yang \cite{Yangsingleauthor} \footnote{for other approaches see \cite{Lam:2018pvp, Mertens:2017mtv, Blommaert:2018oro, Blommaert:2019hjr, Iliesiu:2019xuh}} on the bulk side and compared with RMT predictions for operators satisfying Eigenstate Thermalization Hypothesis (ETH) \cite{ethDeutsch, ethSrednicki} on the boundary side. 

\subsection{Introduction}

The 2d gravity theory we will study consists of the Einstein-Hilbert action + JT gravity action + action from matter. JT gravity on a 2d manifold $M$ has Euclidean action
\begin{equation}
I_{JT}=-\frac{1}{2}\left(\int_M\phi(R+2)+2\int_{\partial M}\phi_b (K-1)\right)
\end{equation}
Classically, the equation of motion fixes the bulk geometry to be AdS$_2$ with $R=-2$ and the action reduces to a Schwarzian action on the boundary \cite{MaldacenaStanfordYang}. In 2d, the Einstein-Hilbert action is purely topological and can be written as 
\begin{equation}
I_{EH}=-\chi S_0
\end{equation}
where $\chi=2-2g-n$ is the Euler character for manifold $M$ with $g$ the genus and $n$ the number of boundaries, and $S_0$ is the zero-temperature bulk entropy which is a constant. The Einstein-Hilbert action then contributes an overall factor $e^{\chi S_0}$ to the partition function. In all of our figures the orange disks represent infinite hyperbolic space (or its quotient) and yellow geometries inside represent the physical Euclidean spacetimes, with wiggly regularized boundaries described by the Schwarzian theory \cite{MaldacenaStanfordYang}.

\begin{figure}[H]
\centering
\includegraphics[width=0.5\textwidth]{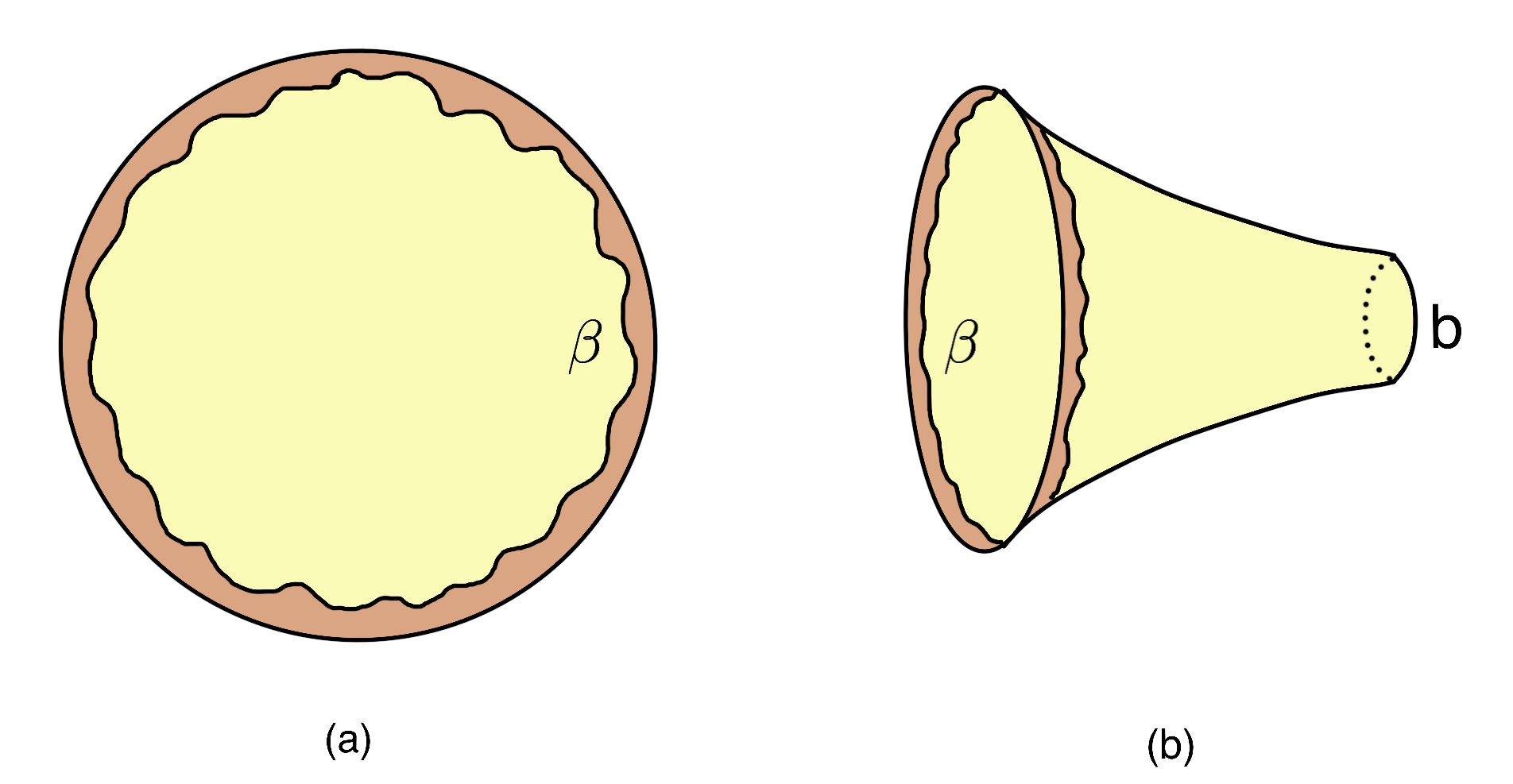}
\caption{(a) disk partition function (b) trumpet partition function}
\label{partitionfunction}
\end{figure}

The two main shapes of Euclidean AdS we consider in this review are a hyperbolic disk which has one asymptotically boundary with renormalized length $\beta$, and a hyperbolic trumpet which has one asymptotic boundary with renormalized length $\beta$ and one geodesic boundary with length $b$ (see figure~\ref{partitionfunction}). That is because a disk is the simplest hyperbolic geometry with one asymptotic boundary and a trumpet can be thought of as a building block of more complicated geometries via attaching a Riemann surface with one geodesic boundary to the geodesic boundary of the trumpet. 

JT path integrals without operator insertions can be computed directly by doing the path integral over the wiggly boundary of the disk and the trumpet explicitly. Disk \cite{BagretsAltlandKamenev16, 9authors, BagretsAltlandKamenev17, StanfordWitten17, SchwarzianBelokurovShavgulidze, SchwarzianMertensTuriaciVerlinde, KitaevSuhBH, Yangsingleauthor, IliesiuPufuVerlindeWang} and trumpet partition functions \cite{StanfordWitten17, Saadsingleauthor} are given respectively by
\begin{equation}
Z_{\text{Disk}}(\beta)=e^{S_0}\frac{e^{\frac{2\pi^2}{\beta}}}{\sqrt{2\pi}\beta^{3/2}}=e^{S_0}\int_0^\infty dE\,\underbrace{\frac{\sinh(2\pi\sqrt{2E})}{2\pi^2}}_{\rho_0(E)}e^{-\beta E}\label{zdisk}
\end{equation}
and 
\begin{equation}
Z_{\text{Trumpet}}(\beta,b)=\frac{e^{-\frac{b^2}{2\beta}}}{\sqrt{2\pi\beta}}=\int_0^\infty dE\,\frac{\cos(b\sqrt{2E})}{\pi\sqrt{2E}}e^{-\beta E}\label{ztrumpet}
\end{equation}
where $\rho_0(E)$ denotes the density of state.

To compute path integrals with operator insertions we need more tools. Before we do that, we should note that there is another way of computing the disk partition function. A disk can be decomposed into two Hartle-Hawking wavefunctions by the following procedure

\begin{align}
Z_{\text{Disk}}(\beta)&=\includegraphics[valign=c,width=0.2\textwidth]{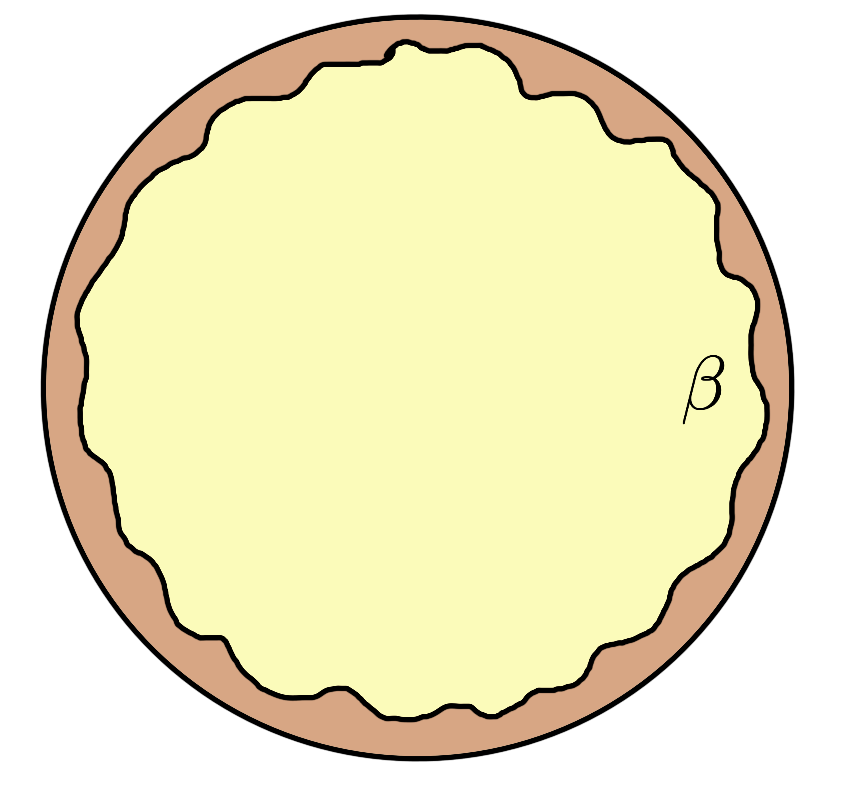}\\
&=\int\,e^\ell d\ell\includegraphics[valign=c,width=0.2\textwidth]{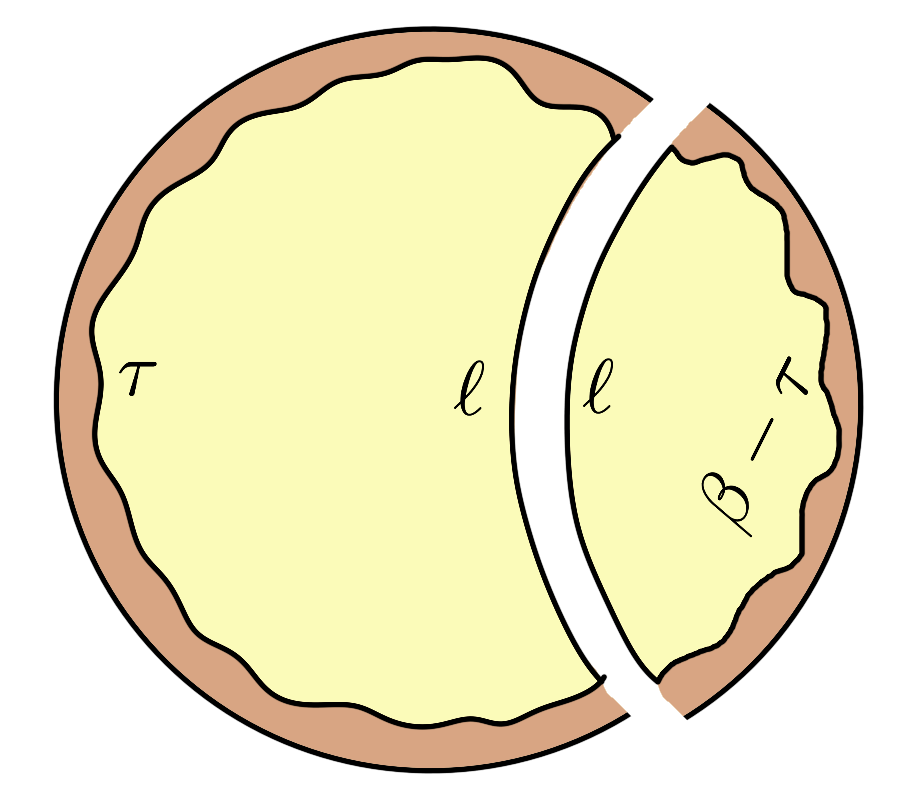}\\
&=e^{S_0}\int\,e^\ell d\ell\,\varphi_{\text{Disk},\tau}(\ell)\varphi_{\text{Disk},\beta-\tau}(\ell)
\end{align}

This decomposition may seem redundant since we already know how to calculate $Z_{\mathrm{disk}}$ but this procedure teaches us how to calculate two-point correlation functions. To do that, we just need another factor of $e^{-\Delta\ell}$ in the integral, which is the QFT two-point correlation function of two boundary operators $V$ of conformal weight $\Delta$ with renormalized geodesic distance $\ell$ apart. Disk contribution to two-point correlation functions at time $t=-i\tau$ is then given by
\begin{align}
\braket{V(t=-i\tau)V(0)}_{\chi=1}&=\includegraphics[valign=c,width=0.2\textwidth]{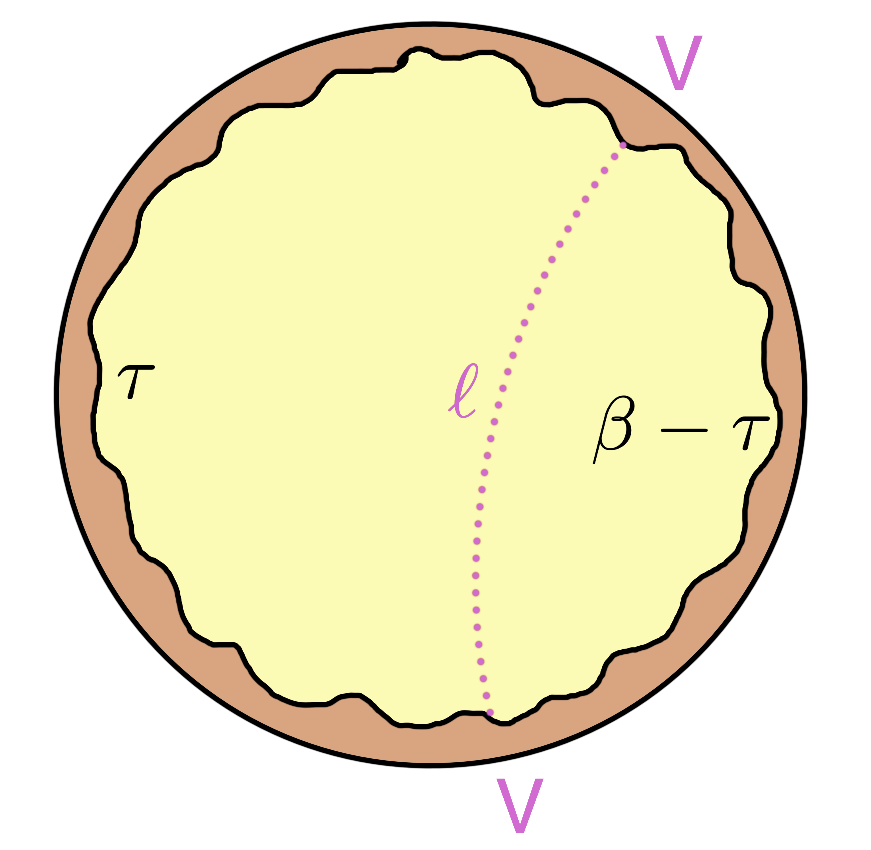}\\
&=\int\,e^\ell d\ell\includegraphics[valign=c,width=0.2\textwidth]{figuresjt/splitcopy.png}e^{-\Delta\ell}\\
&=e^{S_0}\int\,e^\ell d\ell\,\varphi_{\text{Disk},\tau}(\ell)\varphi_{\text{Disk},\beta-\tau}(\ell)e^{-\Delta\ell}
\end{align}
Note that this two-point correlator is not normalized by dividing out the disk partition function which is of order $e^{S_0}$. Hartle-Hawing wavefunctions can be written in a simple closed form by first writing the wavefunctions with fixed energy boundary conditions given by
\begin{equation}
\varphi_E(\ell)=\braket{\ell|E}=4e^{-\ell/2}K_{i\sqrt{8E}}(4e^{-\ell/2})
\end{equation}
where $K$ is a Bessel-K function. Hartle-Hawking wavefunctions i.e. wavefunctions with fixed length boundary condition are given by \cite{Yangsingleauthor,Saadsingleauthor}
\be
\varphi_{\text{Disk},\tau}(\ell)=\int_0^\infty dE\,\rho_0(E)e^{-\tau E}\varphi_E(\ell)\label{hhdisk}
\ee
Now we review two important relations that the Hartle-Hawking wavefunctions satisfy:
\begin{align}
\int_{-\infty}^\infty e^\ell d\ell\,\varphi_E(\ell)\varphi_{E'}(\ell)&=\frac{\delta(E-E')}{\rho_0(E)}\label{hhrelation1}\\
\int_{-\infty}^\infty e^\ell d\ell\,\varphi_E(\ell)\varphi_{E'}(\ell)e^{-\Delta\ell}&=|V_{E,E'}|^2=\frac{\left|\Gamma\left(\Delta+i(\sqrt{2E}+\sqrt{2E'})\right)\Gamma\left(\Delta+i(\sqrt{2E}-\sqrt{2E'})\right)\right|^2}{2^{2\Delta-1}\Gamma(2\Delta)}
\end{align}
In particular using (\ref{hhrelation1}) we can verify 
\be
Z_{\text{Disk}}(\beta)=e^{S_0}\int e^\ell d\ell\,\varphi_{\text{Disk},\tau}(\ell)\varphi_{\text{Disk},\beta-\tau}(\ell)
\ee
by plugging in (\ref{zdisk}, \ref{hhdisk}). In addition to partition functions and Hartle-Hawking states, we review a final and important tool we use: propagators i.e. time evolution operators of Hartle-Hawking wavefunctions such that
\be
\varphi_{\text{Disk},\beta+\beta_1+\beta_2}(\ell)=\int e^{\ell'}d\ell'\,P_{\text{Disk}}(\beta_1,\beta_2,\ell,\ell')\varphi_{\text{Disk},\beta}(\ell')
\ee
we can check \cite{Saadsingleauthor} that the above relations are solved by
\be
P_{\text{Disk}}(\beta_1,\beta_2,\ell,\ell')=\int dE\,\rho_0(E)e^{-(\beta_1+\beta_2)E}\varphi_E(\ell)\varphi_E(\ell')\label{diskpropagator}
\ee

\begin{figure}[h]
\centering
\includegraphics[width=0.5\textwidth]{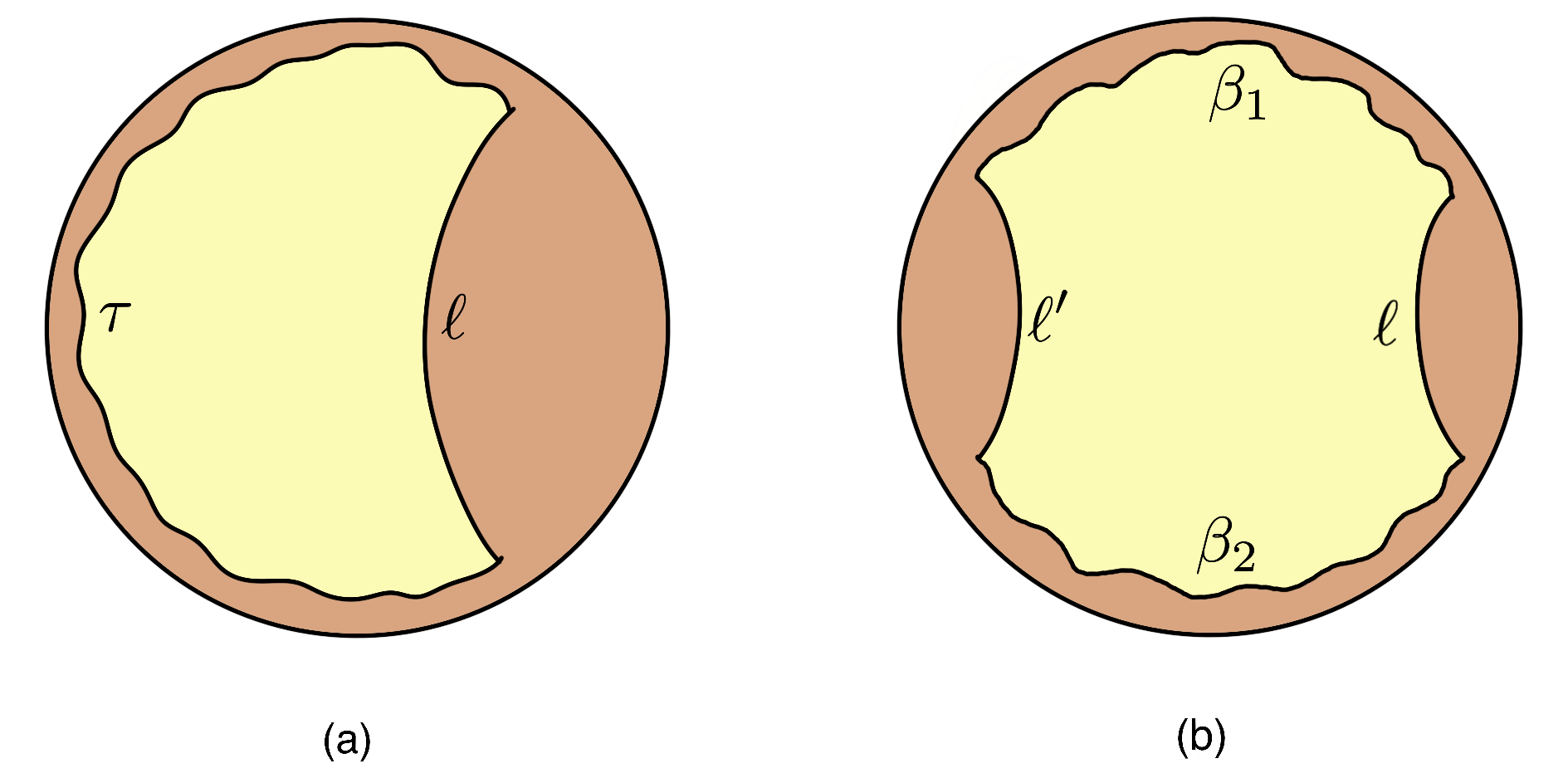}
\caption{(a) Hartle-Hawking wavefunction (b) disk propagator}
\end{figure}

\subsection{Product of two one-point functions}
\label{2donept}
\begin{figure}[H]
\centering
\includegraphics[width=0.5\textwidth]{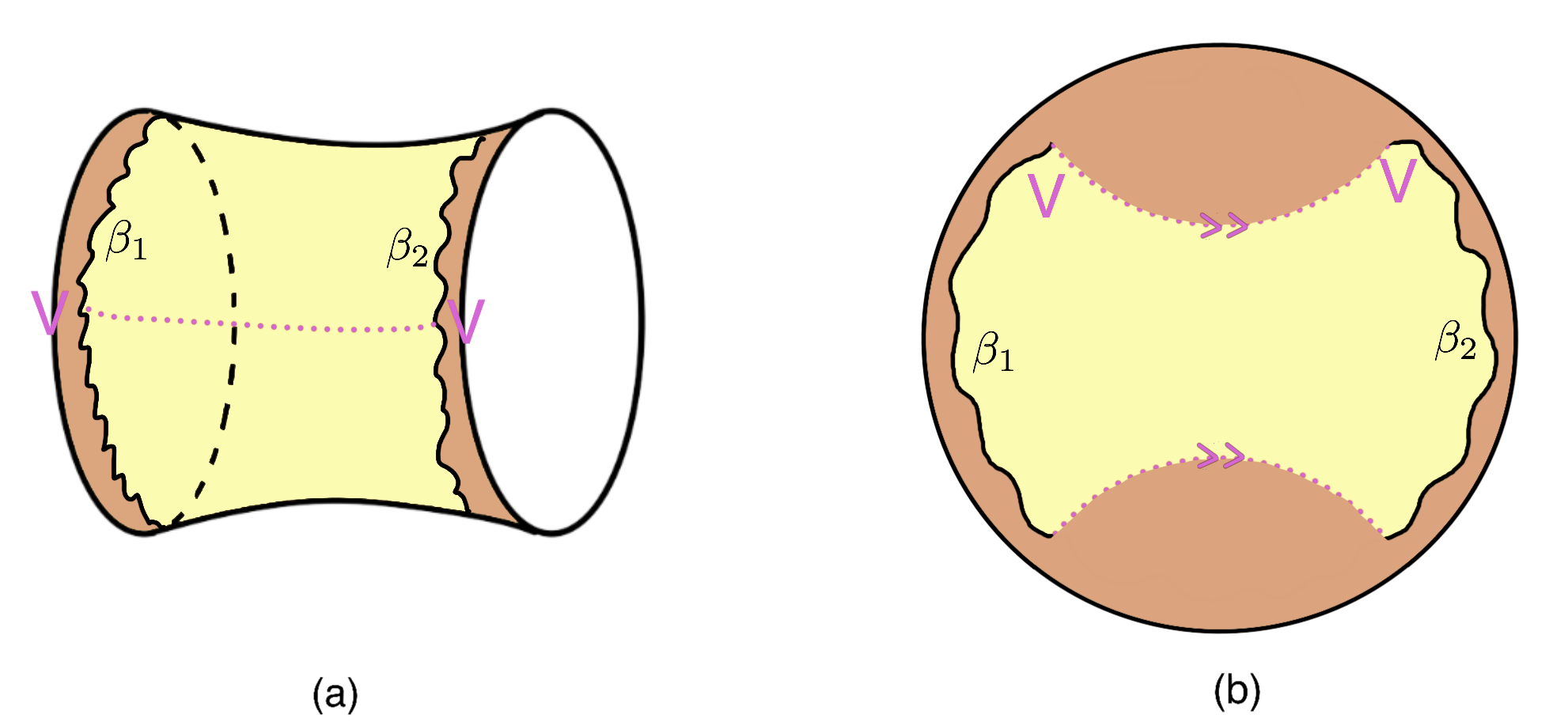}
\caption{(a) a cylinder with one operator insertion on each side (b) embedded on a hyperbolic disk}
\label{2doneptpic}
\end{figure}

In this section, we focus on the product of two one-point functions, which on the gravity side corresponds to a cylinder with one operator insertion on each side as shown in figure \ref{2doneptpic}.

As shown in figure \ref{2doneptpic}(b), this is given by a propagator so
\begin{align}
\braket{V}_{\beta_1}\braket{V}_{\beta_2}&=\int e^\ell d\ell\,P_{\text{Disk}}(\beta_1,\beta_2,\ell,\ell)e^{-\Delta\ell}\\
&=\int dE\,\rho_0(E) e^{-(\beta_1+\beta_2)E}|V_{E,E}|^2\\
&=\int ds\,s\rho_0(s) e^{-(\beta_1+\beta_2)s^2/2}\frac{\Gamma(\Delta\pm i2s)\Gamma(\Delta)^2}{2^{2\Delta-1}\Gamma(2\Delta)}
\end{align}
where $s=\sqrt{2E}$. 

Take the limit $\Delta\rightarrow0$ we get
\begin{align}
\braket{V}_{\beta_1}\braket{V}_{\beta_2}&=4\int ds\,s\rho_0(s) e^{-(\beta_1+\beta_2)s^2/2}\frac{\Gamma(\pm 2is)}{\Delta}\\
&=\frac{4}{\Delta}\int ds\,s\rho_0(s) e^{-(\beta_1+\beta_2)s^2/2}\frac{\pi}{2s\sinh(2\pi s)}
\end{align}
where we have used the fact that when $\Delta\rightarrow0$
\be
\Gamma(\Delta)=\frac{\Gamma(1+\Delta)}{\Delta}=\frac{1}{\Delta}-\gamma+O(\Delta)
\ee
and that
\be
\Gamma(z)\Gamma(-z)=\frac{\Gamma(z)\Gamma(1-z)}{-z}=-\frac{\pi}{z\sin\pi z}
\ee
But now we should recall that the density of states is given by
\be
\rho_0(s)=\frac{\sinh(2\pi s)}{2\pi^2}
\ee
so
\begin{align}
\braket{V}_{\beta_1}\braket{V}_{\beta_2}&=\frac{1}{\pi\Delta}\int ds\,e^{-(\beta_1+\beta_2)s^2/2}\\
&=\frac{1}{\pi\Delta}\sqrt{\frac{\pi}{2(\beta_1+\beta_2)}}\label{limonept}
\end{align}
But on the other hand, we can directly calculate the partition function of a cylinder by using the partition function of trumpet
\begin{align}
Z_{\text{cylinder}}&=\includegraphics[valign=c,width=0.17\textwidth]{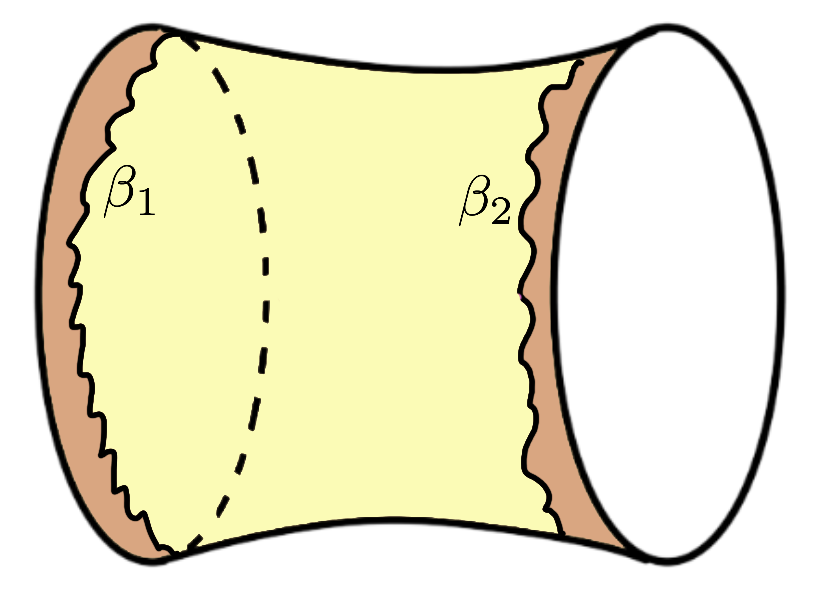}\\
&=\int db\,b\includegraphics[valign=c,width=0.2\textwidth]{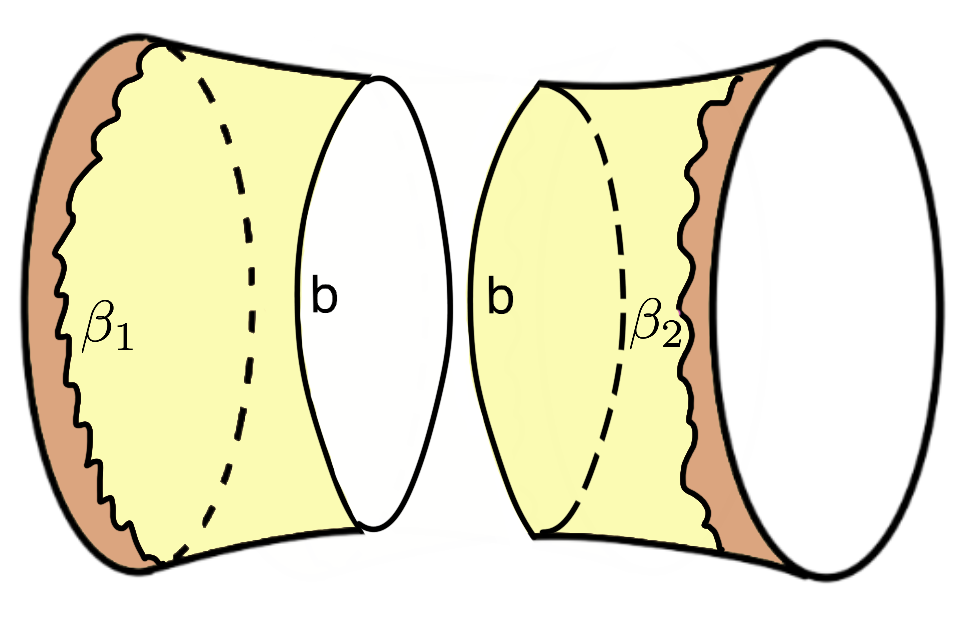}\\
&=\int db\,b\,Z_{\text{Trumpet}}(\beta_1,b)Z_{\text{Trumpet}}(\beta_2,b)\\
&=\frac{\sqrt{\beta_1\beta_2}}{2\pi(\beta_1+\beta_2)}
\end{align}
This disagrees with equation (\ref{limonept}), but we can explain the reason as follows. 

We can also calculate the product of two one-point functions using the trumpet partition function, but when doing that we need to include an additional factor $e^{-\Delta\ell}$. This means that we need to sum over different windings which contributes a factor
\be
\sum_{n=-\infty}^\infty e^{-\Delta b n}\approx\frac{2}{b\Delta}
\ee
Now we can calculate product of two one-point functions again including the winding factor with weight $\Delta\rightarrow0$
\begin{align}
\braket{V}_{\beta_1}\braket{V}_{\beta_2}(\Delta\rightarrow0)&=\int db\,b\,\sum_{n=-\infty}^\infty e^{-\Delta b n}\,Z_{\text{Trumpet}}(\beta_1,b)Z_{\text{Trumpet}}(\beta_2,b)\\
&=\frac{2}{\Delta}\int db\,Z_{\text{Trumpet}}(\beta,b)Z_{\text{Trumpet}}(\beta',b)\\
&=\frac{1}{\pi\Delta}\sqrt{\frac{\pi}{2(\beta_1+\beta_2)}}
\end{align}
This now matches equation (\ref{limonept}) so we have the discrepancy fixed.

\subsection{Product of two two-point functions}
\label{2dtwopt1}
\begin{figure}[H]
\centering
\includegraphics[width=0.5\textwidth]{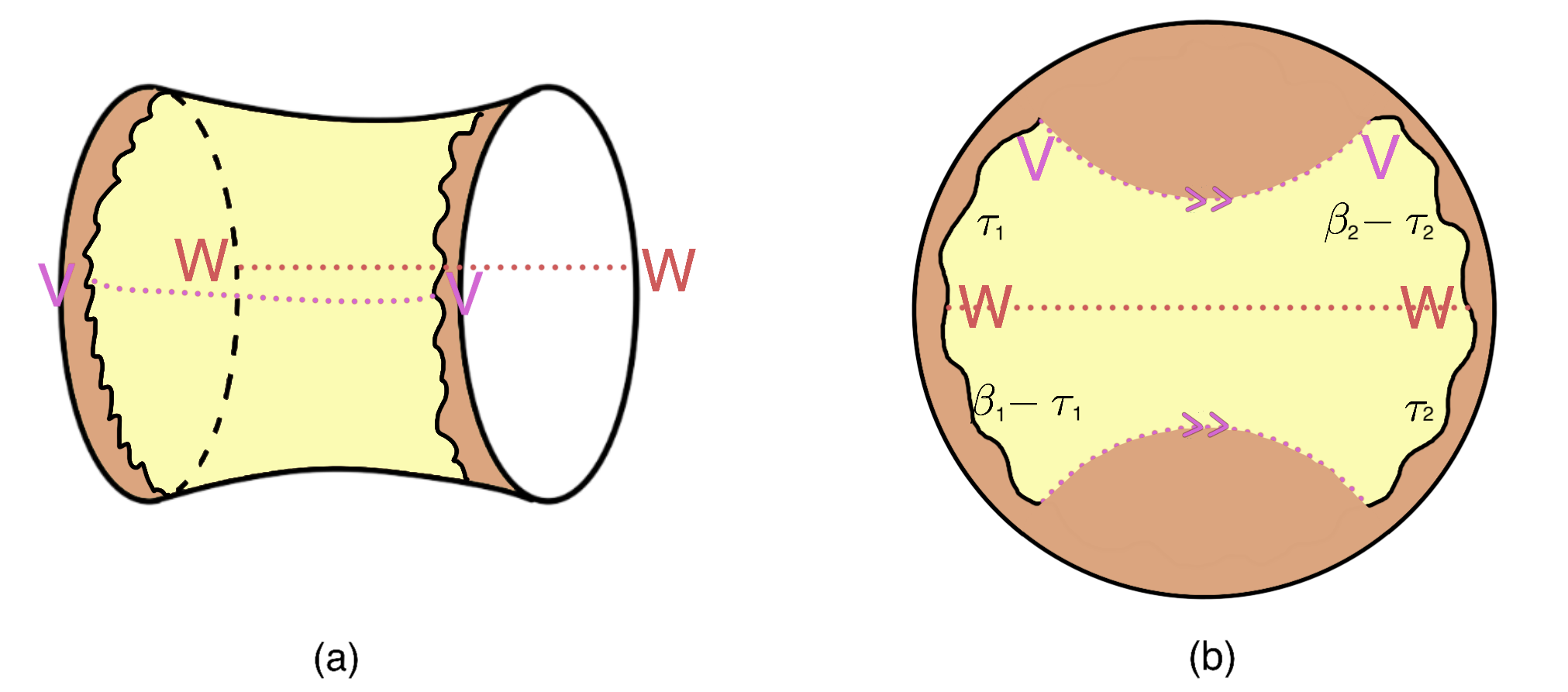}
\caption{(a) a cylinder with two operator insertions on each side (b) embedded on a hyperbolic disk}
\label{2dtwoptpic}
\end{figure}

In this section, we focus on the product of two two-point functions, which on the gravity side corresponds to a cylinder with two operator insertions on each side as shown in figure \ref{2dtwoptpic}.

As shown in figure \ref{2dtwoptpic}(b), this is given by two propagators glued together so
\begin{align}
&\braket{V(\tau_1)W(0)}\braket{V(\beta_2-\tau_2)W(0)}\\
=&\int e^\ell d\ell e^{\ell'}d\ell' P_{\text{Disk}}(\tau_1,\beta_1-\tau_1,\ell,\ell')P_{\text{Disk}}(\beta_2-\tau_2,\tau_2,\ell',\ell)e^{-\Delta_V\ell}e^{-\Delta_W\ell'}\\
=&\int dE dE'\rho_0(E)\rho_0(E')e^{-(\beta_2+\tau_1-\tau_2) E-(\beta_1-\tau_1+\tau_2) E'}|V_{E,E'}|^2|W_{E,E'}|^2
\end{align}
In particular, if we take $V=W$ and $\tau_1=\tau_2$, the above expression becomes
\begin{align}
\braket{V(\tau_1)V(0)}\braket{V(\beta_2-\tau_2)V(0)}&=\int dE dE'\rho_0(E)\rho_0(E')e^{-\beta_2 E-\beta_1 E'}|V_{E,E'}|^4\\
&=\int ds ds' s s'\rho_0(s)\rho_0(s')e^{-\frac{\beta_1s'^2}{2}-\frac{\beta_2 s^2}{2}}\left(\frac{\prod_{\pm_{1,2}}\Gamma\left(\Delta\pm_1 i s\pm_2 i s'\right)}{2^{2\Delta-1}\Gamma(2\Delta)}\right)^2
\end{align}
where $s=\sqrt{2E}$ and $s'=\sqrt{2E'}$.

\subsubsection{crosscap}
\label{2dccapp}
In this section we give a way of calculating the disk with a crosscap contribution to the 2-point correlator by first doubling the crosscap configuration to give a cylinder and then take the square root. For a review of JT gravity framework and 2d crosscap see Appendix \ref{2dcc}. 

We start from the contribution to the averaged product of two 2pt functions. This comes from the cylinder 
\be
\overline{\braket{V(\tau)V(0)}\braket{V(\beta-\tau)V(0)}}=\int dE dE'\underbrace{\frac{\sinh(2\pi \sqrt{2E})}{2\pi^2}}_{\rho_0(E)}\underbrace{\frac{\sinh(2\pi \sqrt{2E'})}{2\pi^2}}_{\rho_0(E')} e^{-\beta(E+E')}|V_{E,E'}|^4\label{2dcylinders}
\ee
where
\be
|V_{E,E'}|^2=\frac{\left|\Gamma\left(\Delta+i(\sqrt{2E}+\sqrt{2E'})\right)\Gamma\left(\Delta+i(\sqrt{2E}-\sqrt{2E'})\right)\right|^2}{2^{2\Delta-1}\Gamma(2\Delta)}
\ee
Let $s=\sqrt{2E}$ and $s'=\sqrt{2E'}$ we can rewrite the above equation as
\be
\overline{\braket{V(\tau)V(0)}\braket{V(\beta-\tau)V(0)}}=\int ds ds' s s'\frac{\sinh(2\pi s)}{2\pi^2}\frac{\sinh(2\pi s')}{2\pi^2}e^{-\frac{\beta}{2}(s^2+s'^2)}\left(\frac{\prod_{\pm_{1,2}}\Gamma\left(\Delta\pm_1 i s\pm_2 i s'\right)}{2^{2\Delta-1}\Gamma(2\Delta)}\right)^2
\ee
Now make a change of variables
\be
s_1+s_2=s_+\quad\quad s_1-s_2=s_-
\ee
Then the above expression becomes
\begin{multline}
\overline{\braket{V(\tau)V(0)}\braket{V(\beta-\tau)V(0)}}\propto\\
\int ds_+ds_-(s_+^2-s_-^2)(e^{2\pi s_+}+ e^{-2\pi s_+}-e^{2\pi s_-}-e^{-2\pi s_-})e^{-\frac{\beta}{4}(s_+^2+s_-^2)}\left(\frac{\prod_{\pm_{1,2}}\Gamma\left(\Delta\pm_1 is_+\right)\Gamma\left(\Delta\pm_2 is_-\right)}{2^{2\Delta-1}\Gamma(2\Delta)}\right)^2
\end{multline}
Using the Stirling approximation 
\be
\prod_{\pm}\Gamma(\Delta\pm is_+)\approx e^{-\pi s_+}s_+^{2\Delta-1}
\ee
So we pick the term $s_+^2$ and $e^{2\pi s_+}$ in the expansion 
\be
\overline{\braket{V(\tau)V(0)}\braket{V(\beta-\tau)V(0)}}\propto\int ds_+ds_- s_+^{4\Delta}e^{-\frac{\beta}{4}(s_+^2+s_-^2)}\left(\frac{\prod_\pm\Gamma\left(\Delta\pm is_-\right)}{2^{2\Delta-1}\Gamma(2\Delta)}\right)^2
\ee
Saddle point of $s_-$ is at $s_-=0$, so (\ref{2dcylinders}) can be approximated by 
\be
\overline{\braket{V(\tau)V(0)}\braket{V(\beta-\tau)V(0)}}\approx \int dE \underbrace{\frac{\sinh(2\pi \sqrt{2E})}{2\pi^2}}_{\rho_0(E)}\underbrace{\frac{\sinh(2\pi \sqrt{2E})}{2\pi^2}}_{\rho_0(E)} e^{-\beta2E}|V_{E,E}|^4
\ee
We should note that here we only care about the saddle point (i.e. only the integrand of the above integral). Thus we get 
\be
\sqrt{\overline{\braket{V(\tau)V(0)}\braket{V(\beta-\tau)V(0)}}}\approx\int dE \underbrace{\frac{\sinh(2\pi \sqrt{2E})}{2\pi^2}}_{\rho_0(E)}e^{-\beta E}|V_{E,E}|^2\label{ccnewmethod}
\ee
In \cite{Yan:2022nod}, crosscap contribution to the 2-point function was calculated directly using propagator
\begin{align}
\braket{V(t=-i\tau)V(0)}_{\text{cc},0}&= \int e^\ell d\ell\,P_{\text{Disk}}(\tau,\beta-\tau,\ell,\ell)e^{-\Delta\ell}\\
&=\includegraphics[valign=c,width=0.25\textwidth]{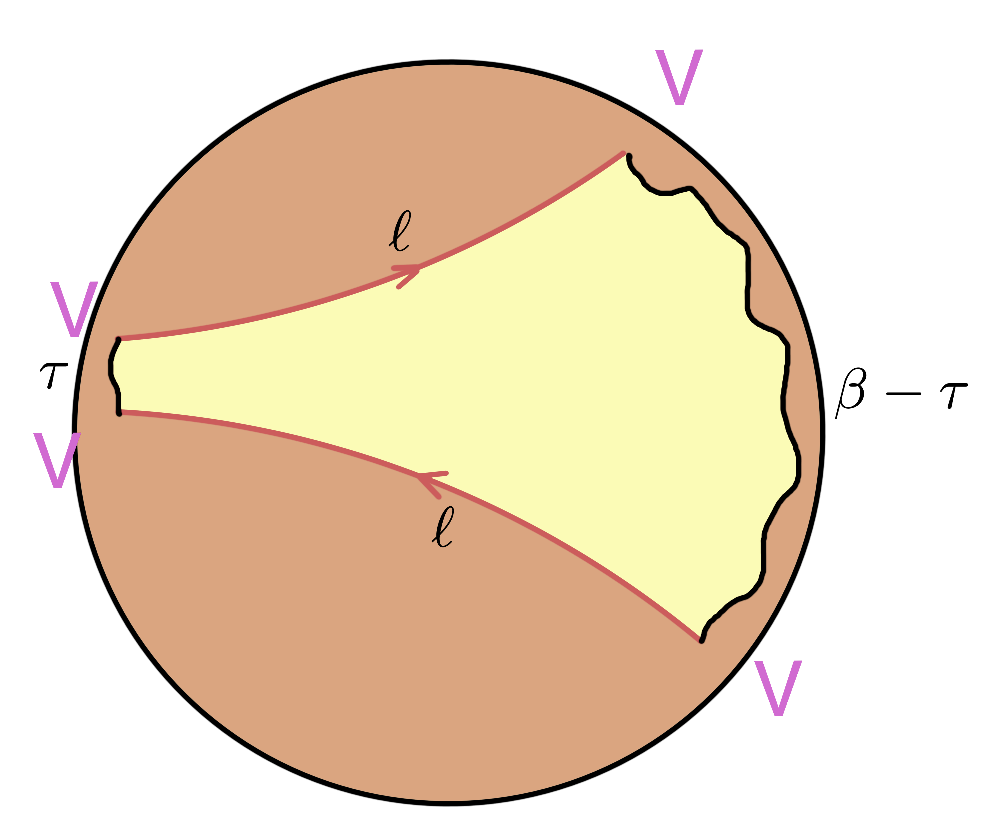}\\
&=\int e^\ell d\ell\,\int dE\,\rho_0(E)e^{-\beta E}\varphi_E(\ell)^2e^{-\Delta\ell}\\
&=\int dE\,\rho_0(E)e^{-\beta E}|V_{E,E}|^2
\end{align}
Therefore (\ref{ccnewmethod}) is exactly the crosscap contribution, so we just showed that
\be
\sqrt{\overline{\braket{V(\tau)V(0)}\braket{V(\beta-\tau)V(0)}}}\approx\overline{\braket{V(t)V(0)}}_{cc}
\ee
\section{Calculation details}
\subsection{Calculating product of two torus one-point functions}
\label{apponept}
We start from the torus partition function
\be
Z(\tau,\bar{\tau})=(q\bar{q})^{-c/24}\mathrm{Tr}(q^{L_0}\bar{q}^{\bar{L}_0})
\ee
where $q=e^{2\pi i\tau}$. Based on this the one-point function can be written as
\begin{align}
\braket{O_1}_{T^2(\tau,\bar{\tau})}&=(q\bar{q})^{-c/24}\mathrm{Tr}(q^{L_0}\bar{q}^{\bar{L}_0}O_1)\\
&=(q\bar{q})^{-c/24}\sum_{h}\braket{h,\bar{h}|O_1|h,\bar{h}}q^{L_0}\bar{q}^{\bar{L}_0}\\
&=\sum_{p}\braket{h_p,\bar{h}_p|O_1|h_p,\bar{h}_p}\sum_{h\in\mathcal{V}_{h_p}}\frac{\braket{h,\bar{h}|O_1|h,\bar{h}}}{\braket{h_p,\bar{h}_p|O_1|h_p,\bar{h}_p}}e^{i\tau(h-c/24)}e^{-i\bar{\tau}(\bar{h}-c/24)}\\
&=\sum_{p}c_{1pp}\sum_{h\in\mathcal{V}_{h_p}}\frac{\braket{h,\bar{h}|O_1|h,\bar{h}}}{\braket{h_p,\bar{h}_p|O_1|h_p,\bar{h}_p}}e^{i\tau(h-c/24)}e^{-i\bar{\tau}(\bar{h}-c/24)}
\end{align}
where we decompose a sum over all operators into a sum over primaries and a sum over all descendents of a primary. Thus, we get an expression for the conformal block
\be
\mathcal{F}_1^{g=1}(h_p;\tau)\overline{\mathcal{F}}_1^{g=1}(\bar{h}_p;\bar{\tau})=\sum_{h\in\mathcal{V}_{h_p}}\frac{\braket{h,\bar{h}|O_1|h,\bar{h}}}{\braket{h_p,\bar{h}_p|O_1|h_p,\bar{h}_p}}e^{i\tau(h-c/24)}e^{-i\bar{\tau}(\bar{h}-c/24)}
\ee
In particular
\be
\mathcal{F}_1^{g=1}(h_p;\tau')\overline{\mathcal{F}}_1^{g=1}(\bar{h}_p;\bar{\tau}')=\overline{\mathcal{F}}_1^{g=1}(h_p;-\tau')\mathcal{F}_1^{g=1}(\bar{h}_p;-\bar{\tau}')
\ee
since complex conjugation gives a minus sign while $\tau\mapsto-\tau$ gives another minus sign, these two minus signs cancel.

Thus the averaged product of two torus one-point functions is given by
\begin{align}
\overline{\braket{O_1}_{T^2(\tau,\bar{\tau})}\braket{O_1}_{T^2(\tau',\bar{\tau}')}}&=\sum_{p,q}\overline{c_{1pp}c_{1qq}}\mathcal{F}_1^{g=1}(h_p;\tau)\overline{\mathcal{F}}_1^{g=1}(\bar{h}_p;\bar{\tau})\mathcal{F}_1^{g=1}(h_q;\tau')\overline{\mathcal{F}}_1^{g=1}(\bar{h}_q;\bar{\tau}')\\
&=2\sum_p\overline{c_{1pp}^2}\mathcal{F}_1^{g=1}(h_p;\tau)\overline{\mathcal{F}}_1^{g=1}(\bar{h}_p;\bar{\tau})\mathcal{F}_1^{g=1}(h_p;\tau')\overline{\mathcal{F}}_1^{g=1}(\bar{h}_p;\bar{\tau}')\\
&=2\sum_p\overline{c_{1pp}^2}\mathcal{F}_1^{g=1}(h_p;\tau)\overline{\mathcal{F}}_1^{g=1}(\bar{h}_p;\bar{\tau})\overline{\mathcal{F}}_1^{g=1}(h_p;-\tau')\mathcal{F}_1^{g=1}(\bar{h}_p;-\bar{\tau}')\\
&\approx2\big|\int dh_p\rho_0(h_p)C_0(h_1,h_p,h_p)\mathcal{F}_1^{g=1}(h_p;\tau)\overline{\mathcal{F}}_1^{g=1}(h_p;-\tau')\big|^2\\
&\approx2\braket{O_1}^L_{T^2(\tau,-\tau')}\braket{O_1}^L_{T^2(-\bar{\tau}',\bar{\tau})}
\end{align}

\subsection{Calculating product of two torus two-point functions}
\label{apptwopt}
The two-point function can be written as
\begin{align}
\braket{O_1(v,\bar{v})O_2(0)}_{T^2(\tau,\bar{\tau})}&=(q\bar{q})^{-c/24}\mathrm{Tr}(q^{L_0}\bar{q}^{\bar{L}_0}O_1O_2)\\
&=(q\bar{q})^{-c/24}\sum_{h,\bar{h},h',\bar{h}'}\braket{h,\bar{h}|e^{i(\tau-v)L_0}e^{-i(\bar{\tau}-\bar{v})\bar{L}_0}O_2e^{ivL_0}e^{-i\bar{v}\bar{L}_0}|h',\bar{h}'}\braket{h',\bar{h}'|O_1|h,\bar{h}}\\
&=\sum_{h,h'}\braket{h,\bar{h}|O_2|h',\bar{h}'}\braket{h',\bar{h}'|O_1|h,\bar{h}}e^{i\tau(h-c/24)}e^{-i\bar{\tau}(\bar{h}-c/24)}e^{iv(h'-h)}e^{-i\bar{v}(\bar{h}-\bar{h}')}\\
&=\sum_{h_p,h_q}\braket{h_p,\bar{h}_p|O_2|h_q,\bar{h}_q}\braket{h_q,\bar{h}_q|O_1|h_p,\bar{h}_p}\nonumber\\
&\times\sum_{h\in\mathcal{V}_{h_p},h'\in\mathcal{V}_{h_q}}\frac{\braket{h,\bar{h}|O_2|h',\bar{h}'}}{\braket{h_p,\bar{h}_p|O_2|h_q,\bar{h}_q}}\frac{\braket{h',\bar{h}'|O_1|h,\bar{h}}}{\braket{h_q,\bar{h}_q|O_1|h_p,\bar{h}_p}}e^{i\tau(h-c/24)}e^{-i\bar{\tau}(\bar{h}-c/24)}e^{iv(h'-h)}e^{-i\bar{v}(\bar{h}-\bar{h}')}\\
&=\sum_{h_p,h_q}c_{1pq}c_{2pq}\mathcal{F}_{12}^{g=1}(h_p,h_q;\tau,v)\overline{\mathcal{F}}_{12}^{g=1}(\bar{h}_p,\bar{h}_q;\bar{\tau},\bar{v})
\end{align}
Comparing these two results, we get an expression for the conformal block
\begin{multline}
\mathcal{F}_{12}^{g=1}(h_p,h_q;\tau,v)\overline{\mathcal{F}}_{12}^{g=1}(\bar{h}_p,\bar{h}_q;\bar{\tau},\bar{v})\\
=\sum_{h\in\mathcal{V}_{h_p},h'\in\mathcal{V}_{h_q}}\frac{\braket{h,\bar{h}|O_2|h',\bar{h}'}}{\braket{h_p,\bar{h}_p|O_2|h_q,\bar{h}_q}}\frac{\braket{h',\bar{h}'|O_1|h,\bar{h}}}{\braket{h_q,\bar{h}_q|O_1|h_p,\bar{h}_p}}e^{i\tau(h-c/24)}e^{-i\bar{\tau}(\bar{h}-c/24)}e^{iv(h'-h)}e^{-i\bar{v}(\bar{h}-\bar{h}')}
\end{multline}
In particular we have
\be
\mathcal{F}_{12}^{g=1}(h_p,h_q;\tau',v')\overline{\mathcal{F}}_{12}^{g=1}(\bar{h}_p,\bar{h}_q;\bar{\tau}',\bar{v}')=\overline{\mathcal{F}}_{12}^{g=1}(h_p,h_q;-\tau',-v')\mathcal{F}_{12}^{g=1}(\bar{h}_p,\bar{h}_q;-\bar{\tau}',-\bar{v}')
\ee
Thus the averaged product of two torus two-point functions is given by
\begin{align}
&\overline{\braket{O_1(v,\bar{v})O_2(0)}_{T^2(\tau,\bar{\tau})}\braket{O_1(v',\bar{v}')O_2(0)}_{T^2(\tau',\bar{\tau}')}}\nonumber\\
=&\sum_{p,q,r,s}\overline{c_{1pq}c_{2pq}c_{1rs}c_{2rs}}\mathcal{F}_{12}^{g=1}(h_p,h_q;\tau,v)\overline{\mathcal{F}}_{12}^{g=1}(\bar{h}_p,\bar{h}_q;\bar{\tau},\bar{v})\mathcal{F}_{12}^{g=1}(h_r,h_s;\tau',v')\overline{\mathcal{F}}_{12}^{g=1}(\bar{h}_r,\bar{h}_s;\bar{\tau}',\bar{v}')\\
=&2\sum_{p,q}\overline{|c_{1pq}^2|}\overline{|c_{2pq}^2|}\mathcal{F}_{12}^{g=1}(h_p,h_q;\tau,v)\overline{\mathcal{F}}_{12}^{g=1}(\bar{h}_p,\bar{h}_q;\bar{\tau},\bar{v})\mathcal{F}_{12}^{g=1}(h_p,h_q;\tau',v')\overline{\mathcal{F}}_{12}^{g=1}(\bar{h}_p,\bar{h}_q;\bar{\tau}',\bar{v}')\\
=&2\sum_{p,q}\overline{|c_{1pq}^2|}\overline{|c_{2pq}^2|}\mathcal{F}_{12}^{g=1}(h_p,h_q;\tau,v)\overline{\mathcal{F}}_{12}^{g=1}(\bar{h}_p,\bar{h}_q;\bar{\tau},\bar{v})\overline{\mathcal{F}}_{12}^{g=1}(h_p,h_q;-\tau',-v')\mathcal{F}_{12}^{g=1}(\bar{h}_p,\bar{h}_q;-\bar{\tau}',-\bar{v}')\\
\approx&2\big|\int dh_p dh_q\rho_0(h_p)\rho_0(h_q)C_0(h_1,h_p,h_q)C_0(h_2,h_p,h_q)\mathcal{F}_{12}^{g=1}(h_p,h_q;\tau,v)\overline{\mathcal{F}}_{12}^{g=1}(h_p,h_q;-\tau',-v')\big|^2\\
\approx& 2\braket{O_1(v,-v')O_2(0)}^L_{T^2(\tau,-\tau')}\braket{O_1(-\bar{v}',\bar{v})O_2(0)}^L_{T^2(-\bar{\tau}',\bar{\tau})}
\end{align}

\subsection{Averaged product of two torus one-point function in $\Delta\rightarrow0$ limit}
\label{appdetailc0}

Let us begin with our formula of the averaged product of two one-point functions
\be
\overline{\braket{O_1}_{T^2(\tau,\bar{\tau})}\braket{O_1}_{T^2(\tau',\bar{\tau}')}}\approx2\big|\int dh_p\rho_0(h_p)C_0(h_1,h_p,h_p)\mathcal{F}_1^{g=1}(h_p;\tau)\overline{\mathcal{F}}_1^{g=1}(h_p;-\tau')\big|^2
\ee
where $C_0$ is given by 
\be
C_0(P_1,P_2,P_3)=\frac{\Gamma_b(2Q)}{\sqrt{2}\Gamma_b(Q)^3}\frac{\Gamma_b(\frac{Q}{2}\pm iP_1\pm iP_2 \pm iP_3)}{\prod_{k=1}^3\Gamma_b(Q\pm 2iP_k)}
\ee
Now we take the weight of $O_1$ to zero i.e. $\Delta=2h=2\alpha(Q-\alpha)\rightarrow0$. Without loss of generality we take $\alpha=\epsilon\rightarrow0$, so $\Delta=2\epsilon Q$. More specifically $iP_1=-\frac{Q}{2}+\epsilon$ so
\be
C_0(h_1,h_p,h_p)=C_0(i(\frac{Q}{2}-\epsilon),P,P)=\frac{\Gamma_b(\epsilon)^2\Gamma_b(\pm2iP)}{\sqrt{2}\Gamma_b(Q)\Gamma_b(2\epsilon)\Gamma_b(Q\pm 2iP)}
\ee
Remember the definition of $\Gamma_b$ in terms of double gamma function
\be
\Gamma_b(P)=\frac{\Gamma_2(P|b,b^{-1})}{\Gamma_2(\frac{Q}{2}|b,b^{-1})}
\ee
where
\be
\Gamma_2(w|a_1,a_2)=\frac{e^{\lambda_1w+\lambda_2w^2}}{w}\prod_{(n_1,n_2)\in\N^2,(n_1,n_2)\neq(0,0)}\frac{e^{\frac{w}{n_1a_1+n_2a_2}-\frac{1}{2}\frac{w^2}{(n_1a_1+n_2a_2)^2}}}{1+\frac{w}{n_1a_1+n_2a_2}}
\ee
so as we take $\epsilon\rightarrow0$
\be
\Gamma_2(\epsilon|b,b^{-1})=\frac{1}{\epsilon}
\ee
and
\be
C_0(h_1,h_p,h_p)=\frac{\sqrt{2}}{\epsilon\Gamma_2(Q|b,b^{-1})}\frac{\Gamma_b(\pm2iP)}{\Gamma_b(Q\pm2iP)}
\ee
Now observe that $\Gamma_b$ satisfies the following relations
\be
\Gamma_b(w+b)=\sqrt{2\pi}\frac{b^{bw-\frac{1}{2}}}{\Gamma(bw)}\Gamma_b(w)\quad\quad\Gamma_b(w+b^{-1})=\sqrt{2\pi}\frac{b^{-b^{-1}w+\frac{1}{2}}}{\Gamma(b^{-1}w)}\Gamma_b(w)
\ee
so correspondingly
\be
\Gamma_b(Q+2iP)=\frac{2\pi b^{1+i2(b-b^{-1})P}\Gamma_b(2iP)}{\Gamma(1+2ibP)\Gamma(2ib^{-1}P)}\quad\quad\Gamma_b(Q-2iP)=\frac{2\pi b^{1-i2(b-b^{-1})P}\Gamma_b(-2iP)}{\Gamma(1-2ibP)\Gamma(-2ib^{-1}P)}
\ee
so
\be
\frac{\Gamma_b(\pm2iP)}{\Gamma_b(Q\pm2iP)}=\frac{\Gamma(1\pm 2ibP)\Gamma(\pm 2ib^{-1}P)}{(2\pi)^2b^2}=\frac{1}{4\sinh(2\pi b P)\sinh(2\pi b^{-1}P)}
\ee
where we have used 
\begin{align}
\Gamma(1+z)\Gamma(1-z)&=z\Gamma(z)\Gamma(1-z)=\frac{z\pi}{\sin\pi z}\\
\Gamma(z)\Gamma(-z)&=\frac{\Gamma(z)\Gamma(1-z)}{-z}=-\frac{\pi}{z\sin\pi z}
\end{align}
so
\begin{align}
C_0(h_1,h_p,h_p)&=\frac{1}{16\epsilon\Gamma_2(Q|b,b^{-1})\sinh^2(2\pi b P)\sinh^2(2\pi b^{-1}P)}\\
&=\frac{Q}{8\Delta\Gamma_2(Q|b,b^{-1})\sinh^2(2\pi b P)\sinh^2(2\pi b^{-1}P)}
\end{align}
From \cite{Collier:2019weq} equation (2.14) we know that
\be
\rho(P,\bar{P})=\rho_0(P)\rho_0(\bar{P})=|4\sqrt{2}\sinh(2\pi bP)\sinh(2\pi b^{-1}P)|^2
\ee
so now we can simplify the product of two one-point functions
\be
\overline{\braket{O_1}_{T^2(\tau,\bar{\tau})}\braket{O_1}_{T^2(\tau',\bar{\tau}')}}\approx\frac{32Q^2}{\Delta^2\Gamma_2(Q|b,b^{-1})^2}\left|\int dP\,\mathcal{F}_1^{g=1}(h_p;\tau)\overline{\mathcal{F}}_1^{g=1}(h_p;-\tau')\right|^2
\ee
Now we evaluate $\Gamma_2(Q|b,b^{-1})$ using the formula
\be
\Gamma_N(w|a_1,\ldots,a_N)=\Gamma_{N-1}(w|a_1,\ldots,a_{N-1})\Gamma_N(w+a_N|a_1,\ldots,a_N)
\ee
For $N=2$
\be
\Gamma_2(w|a_1,a_2)=\Gamma_1(w|a_1)\Gamma_2(w+a_2|a_1,a_2)
\ee
Then
\begin{align}
\Gamma_2(\epsilon+b+\frac{1}{b})&=\frac{\Gamma_2(\epsilon+b|b,\frac{1}{b})}{\Gamma(\epsilon+b|b)}=\frac{\Gamma_2(\epsilon|b,\frac{1}{b})}{\Gamma(b|b)\Gamma(\epsilon|\frac{1}{b})}
\end{align}
Using
\be
\Gamma(w|a)=\frac{a^{a^{-1}w-\frac{1}{2}}}{\sqrt{2\pi}}\Gamma(a^{-1}w)
\ee
\begin{align}
\Gamma(b|b)&=\sqrt{\frac{b}{2\pi}}\Gamma(1)=\sqrt{\frac{b}{2\pi}}\\
\Gamma(\epsilon|\frac{1}{b})&=\sqrt{\frac{b}{2\pi}}\Gamma(b\epsilon)
\end{align}
Recall that as $w\rightarrow0$
\be
\Gamma(w)=\frac{1}{w}+O(w^0)\quad\quad \Gamma_2(w|b,b^{-1})=\frac{1}{w}+O(w^0)
\ee
so
\be
\Gamma_2(\epsilon+b+\frac{1}{b}|b,\frac{1}{b})=\frac{\frac{1}{\epsilon}}{\frac{b}{2\pi}\frac{1}{b\epsilon}}=2\pi
\ee
Therefore,
\be
\overline{\braket{O_1}_{T^2(\tau,\bar{\tau})}\braket{O_1}_{T^2(\tau',\bar{\tau}')}}\approx\frac{8Q^2}{\Delta^2\pi^2}\left|\int dP\,\mathcal{F}_\mathbb{1}^{g=1}(h_p;\tau)\overline{\mathcal{F}}_\mathbb{1}^{g=1}(h_p;-\tau')\right|^2
\ee

\subsection{Saddle-point approximation in 3d}
\label{saddle}

Saddle-point approximation was proven to work in an analogous situation in 2d (for details see Appendix \ref{2dtwopt1}). In 2d, we calculate the averaged product of two two-point functions and take the saddle-point approximation. We compare this result with contribution to two-point function from disk with crosscap (see Appendix \ref{2dccapp}) and results are the same. We then assume here that saddle-point approximation can be generalized to 3d even though we do not have a direct exact calculation in this case.

We already know from calculations in previous sections
\begin{multline}
\overline{\braket{O_1(v,\bar{v})O_2(0)}_{T^2(\tau,\bar{\tau})}\braket{O_1(v',\bar{v}')O_2(0)}_{T^2(\tau',\bar{\tau}')}}=\\
2\big|\int dh_2 dh_3\rho_0(h_2)\rho_0(h_3)C_0(h_1,h_2,h_3)C_0(h_1,h_2,h_3)\mathcal{F}_{11}^{g=1}(h_2,h_3;\tau,v)\mathcal{F}_{11}^{g=1}(h_2,h_3;\tau',v')\big|^2
\end{multline}
where
\be
\tau=i\beta\quad\tau'=i\beta\quad v=i\frac{\beta}{2}-t\quad v'=i\frac{\beta}{2}+t
\ee
Recall that
\be
\rho_0(h)\approx \exp\left(2\pi\sqrt{\frac{c}{6}(h-\frac{c}{24})}\right)
\ee
and recall that the functions $C_0$ are given by
\be
C_0(P_1,P_2,P_3)=\frac{\Gamma_b(2Q)}{\sqrt{2}\Gamma_b(Q)^3}\frac{\prod_{\pm_{1,2,3}}\Gamma_b(\frac{Q}{2}\pm_1 iP_1\pm_2 iP_2\pm_3iP_3)}{\prod_{k=1}^3\Gamma_b(Q+2iP_k)\Gamma_b(Q-2iP_k)}
\ee
so we can again define 
\be
P_+=P_2+P_3\quad\quad P_-=P_2-P_3
\ee
then the saddle points are at $P_+=0$ and $P_-=0$, i.e. $h_2=h_3=h_p$. In particular, we should note that for the particular values of $\tau,\tau',v,v'$ that we are considering
\be
\mathcal{F}_{11}^{g=1}(h_p,h_p;\tau,v)=\mathcal{F}_{11}^{g=1}(h_p,h_p;\tau',v')
\ee
Therefore, we have 
\be
\sqrt{\overline{\braket{O_1(v,\bar{v})O_2(0)}_{T^2(\tau,\bar{\tau})}\braket{O_1(v',\bar{v}')O_2(0)}_{T^2(\tau',\bar{\tau}')}}}\approx|\int dh_p\rho_0(h_p)C_0(h_1,h_p,h_p)\mathcal{F}_{11}^{g=1}(h_p,h_p;\tau,v)|^2
\ee

\bibliography{references}

\bibliographystyle{utphys}

\end{document}